\documentclass[twocolumn,tighten,time]{aastex631}
\usepackage[fleqn]{amsmath}
\usepackage{graphicx}	
\pdfoutput=1 
\usepackage{epsfig}
\usepackage{subfigure}
\usepackage{multirow}

\newcommand{\ii}{_i}
\newcommand{\der}{\mathrm{d}}
\newcommand{\R}{R_{\rm f}}
\newcommand{\pk}{_{\rm pk}}

\newcommand{\p}{_{\rm p}} 

\newcommand{\vir}{_{\rm vir}}

\newcommand{\f}{_{\rm sf}}

\newcommand{\res}{_{\rm min}} 
\newcommand{\acc}{^{\rm acc}}
\newcommand{\dc}{_{\rm dDM}}

\newcommand{\W}{_{\rm W}} 
\newcommand{\E}{_{\rm E}} 
\newcommand{\X}{_{\rm WDM}}
\newcommand{\s}{_{\rm fs}} 

\newcommand{\F}{^{\rm th}}
\newcommand{\G}{}

\newcommand{\m}{_{-1}} 
\newcommand{\el}{^{\rm iso}}

\newcommand{\nest}{^{\rm nst}}
\newcommand{\fnest}{^{\rm d\,nst}}

\newcommand{\clM}{M_{\rm s}}

\newcommand{\tr}{^{\rm tr}} 
\newcommand{\st}{_{\rm rel}} 
\newcommand{\str}{^{\rm stp}} 

\newcommand{\maxi}{_{\rm max}}
\newcommand{\wR}{R_{\rm s}} 

\newcommand{\h}{_{\rm h}}

\newcommand{\jpk}{_{j}}
\newcommand{\jp}{_{\rm p}j}
\newcommand{\po}{_{\rm p1}}
\newcommand{\ptt}{_{\rm p2}}
\newcommand{\pz}{_{\rm p3}}
\newcommand{\ho}{_{\rm h1}}
\newcommand{\htt}{_{\rm h2}}
\newcommand{\hz}{_{\rm h3}}
\newcommand{\opk}{_{\rm 1}}
\newcommand{\tpk}{_{\rm 2}}
\newcommand{\zpk}{_{\rm 3}}

\newcommand{\beq}{\begin{equation}} 
\newcommand{\eeq}{\end{equation}}
\newcommand{\beqa}{\begin{eqnarray}}
\newcommand{\eeqa}{\end{eqnarray}} 
\newcommand{\derpr}{\partial_{\rm r}}
\newcommand{\sphc}{(r,\theta,\varphi)}

\newcommand{\ep}{\epsilon}
\newcommand{\es}{{\varepsilon}}
\newcommand{\eps}{e}
\newcommand{\esp}{\tilde e}

\newcommand{\cc}{_{\rm c}}

\newcommand{\crit}{_{\rm crit}}

\newcommand{\ti}{t_{\rm i}}

\newcommand{\M}{_{\rm M}}

\newcommand{\modotb}{M$_\odot$\ } 
\newcommand{\modot}{M$_\odot$}
\newcommand{\lav}{\langle}
\newcommand{\rav}{\rangle}
\newcommand{\rad}{_{\rm r}} 
\newcommand{\tang}{_{\rm t}}

\begin{document}

\shorttitle{Culminating the Peak CUSP}
\shortauthors{Salvador-Sol\'e et al.}

\title{CULMINATING THE PEAK CUSP TO DESCRY THE DARK SIDE OF HALOS}

\author{Eduard Salvador-Sol\'e and Alberto Manrique}
\affiliation{Institut de Ci\`encies del Cosmos. Universitat de Barcelona, E-08028 Barcelona, Spain}

\email{e.salvador@ub.edu}



\begin{abstract}
The {\it ConflUent System of Peak trajectories} (CUSP) is a rigorous formalism in the framework of the peak theory that allows one to derive from first principles andno free parameters the typical halo properties from the statistics of peaks in the filtered Gaussian random field of density perturbations. The predicted halo mass function, spherically averaged density, velocity dispersion, velocity anisotropy, ellipticity, prolateness and potential profiles, as well as the abundance and number density profiles of accreted and stripped subhalos and diffuse dark matter accurately recover the results of cosmological $N$-body simulations. CUSP is thus a powerful tool for the calculation, in any desired hierarchical cosmology with Gaussian perturbations, of halo properties beyond the mass, redshift and radial ranges covered by simulations. More importantly, CUSP unravels the origin of the characteristic features of those properties. In the present Paper we culminate its construction. We show that all halo properties but those related with subhalo stripping are independent of the assembly history of those objects, and that the Gaussian is the only smoothing window able to find the finite collapsing patches while properly accounting for the entropy increase produced in major mergers.
\end{abstract}

\keywords{methods: analytic --- cosmology: theory, dark matter --- dark matter: halos --- galaxies: halos, substructure}


\section{Introduction}\label{intro}

One fundamental limitation for the theory of structure formation is the lack of an exact analytic treatment in non-linear regime. Indeed, such a treatment is available for monolithic spherical collapse. But the typical seeds of dark matter (DM) halos are ellipsoidal, and so is also their collapse. Moreover, halos suffer major mergers, so their collapse is rather lumpy. This is the reason that halo properties have been traditionally studied by means of numerical simulations (see e.g. the review by \citealt{FW12}).

But simulations are not without problems. They are very CPU-time consuming, and hence, only deal with moderately large numbers of particles. As a consequence, their dynamic range and spatial resolution are limited. In addition, their analysis involves complex selection procedures similar to those used in observations which may bias the results. Last but not least, even though they are very useful to determine halo properties, they are not well-suited to explain the origin of those properties.

To shed light on that `dark side' of halos, structure formation has also been studied by analytic means. \citet{PS} used the top-hat spherical collapse approximation to infer the halo mass function (MF) from the statistics of perturbations in the initial Gaussian random density field. This approach was refined by correcting for cloud-in-cloud configurations \citep{BCEK}, extended to calculate conditional MFs and merger rates \citep{B91,LC93}, and modified to account for ellipsoidal collapse \citep{M95,Sea01,ST02}. Some authors \citep{CLM89,PH90,AJ90,bm,ESP} went a step further and took into account that halos evolve from maxima (peaks) in the filtered initial density field (\citealt{D70,BBKS}, hereafter BBKS). One particularly elaborate treatment along this line was the so-called {\it ConflUent System of Peak trajectories} (CUSP) formalism developed by \citet{MSS95,MSS96}, which assumed the existence of a halo-peak correspondence, calibrated by means of the results of simulations.

Other authors concentrated in deriving the density profile for halos that result from the spherical collapse of the homogeneous mass distribution around a density perturbation, the so-called secondary-infall model (\citealt{GG72,FG84,Ber85}), or from the spherical collapse of the perturbation itself \citep{ARea98,metal03,smgh07}. This approach was also pursued within the peak theory \citep{DPea00,As04,McM06,Sea12a,Sea12b}.

However, the analytic approach faces the following apparently insurmountable fundamental difficulties (Ds):

\begin{itemize}

\item D i) Even though halos seem to arise from the collapse of patches with negative total energy corresponding to peaks in the smothed initial density field \citep{HP13}, the relation between halos and protohalos in simulations is not the expected one (\citealt{LP}).

\item D ii) Moreover, there seems to be no one-to-one correspondence between halos and peaks. Peaks of a given density contrast are {\it overcrowded} (\citealt{AJ90}), and halo seeds often split in a few disjoint nodes \citep{Pea}.

\item D iii) The edge of a virialized halo\footnote{The virial relation we refer to throughout this Paper includes the external pressure term, so by virialized halos we simply mean halos in (quasi)equilibrium.} is a fuzzy concept, there being many possible halo mass definitions (e.g. \citealt{Dea85}).

\item D iv) There is no clear argument in favor of any particular smoothing window (e.g. \citealt{B88}) and the mass encompassed by a window of a given scale is unknown.\footnote{It is only known for the top-hat window, but there is no obvious relation between the {\it mass}-dependent 0-th order spectral moments for different windows.}

\item D v) Peaks are triaxial (\citealt{D70}) and the time of their ellipsoidal collapse, along one axis first (pancakes), then another axis (filaments) and the third axis in the end \citep{LMS65,Z70} is unkown.

\item D vi) As DM is collisionless, monolithic collapse is followed by shell-crossing \citep{H64}, with no analytic treatment not even assuming spherical collapse.

\item D vii) In addition, the real collapse of halos is lumpy, i.e. they suffer major mergers producing a violent relaxation with no analytic treatment and a poorly known final state \citep{LB67,S78}.

\item D viii) The material assembled in halos is a mixture of diffuse DM (dDM) and other halos \citep{AW}, which become subhalos tidally stripped by the host potential well. The situation is thus very complex (e.g. \citealt{Gea98}).
 
\item D ix) In addition, massive subhalos suffer dynamical friction and eventually merge at the halo center, whereas the exact treatment of dynamical friction is only available for infinite homogeneous systems \citep{C43}.

\end{itemize}

Yet, using CUSP, \citet{Sea12a,Sea12b} were able to derive not only the typical density profile, but also the velocity dispersion and anisotropy profiles and even the ellipticity and prolateness profiles of halos in CDM cosmologies (see also \citealt{Vea12} for WDM cosmologies). \citet{Jea14a,Jea14b} formally proved the basic hypothesis of CUSP that there is a one-to-one correspondence between halos and peaks and rederived the halo MF and density profile from first principles and with no single free parameter. Lastly, \citet{Sea19a,Sea19b,Sea19c} have recently derived the properties of halo substructure. 

All CUSP predictions are in very good agreement with the results of simulations, so this formalism can be used to extend the halo properties beyond the radius, mass, and redshift ranges and cosmologies covered by simulations. More importantly, CUSP unravels the origin of all those properties and their characteristic features. Yet CUSP has not attracted much attention possibly because of its non-linear construction over many years, the lengthy mathematical developments included in its development, and the fact that it has never been clearly shown how CUSP solves or avoids the above mentioned Ds for an accurate analytic treatment of structure formation so that the prejudice that this is not possible remains. 

More importantly, CUSP has two weak points. First, the derivation of all halo properties is made assuming monolithic collapse, while real halos suffer major mergers. The fact that CUSP recovers, nonetheless, the results of simulations seems to imply that the inner properties of halos do not depend on their assembly history perhaps because in major mergers halos lose the memory of their past history \citep{Sea12a}. Some results of simulations support, indeed, this idea \citep{Hea99,Hea06,WW09,Bea12}, but others point to the opposite conclusion \citep{Gea01,Gea02,ST04,FM09,FM10,Hea09}. Second, the properties of peaks from which all halo properties follow depend on the particular smoothing window used to filter the initial density field. CUSP employs the Gaussian window for practical reasons (it greatly simplifies the calculations; BBKS), but the reason why this particular window leads to such good results is poorly understood.

In this paper we address these issues. We prove that the properties of halos do not depend on their assembly history (except for the properties related to subhalo stripping), and show that the use of a Gaussian filter is mandatory for the filtering of the initial density field to properly trace DM clustering. With these results we culminate the construction of CUSP.

Taking advantage of the opportunity, we provide a compact orderly review of CUSP skipping all detailed mathematical developments and showing, instead, how CUSP solves or circumvents the above mentioned Ds and clarifies some intriguing questions (Qs) risen by simulations, listed in the end of the Paper. To this end, each time one D is solved or one Q is answered it is indicated in parentheses.

The layout of the Paper is as follows. In Section \ref{peak-halo} we establish the halo-peak correspondence at the base of CUSP. In Section \ref{MF} we derive the halo and subhalo MFs from peak counts. The inner properties of halo seeds are derived in Section \ref{protohalo} and those of halos following from them are given in Section \ref{accretion}. The role of major mergers and the Gaussian window is addressed in Section \ref{mergers}. The implications of our results are discussed in Section \ref{summ}. Throughout the Paper we provide some Figures in order to illustrate the goodness of the CUSP predictions, calculated for the LCDM {\it WMAP7} cosmology and using the BBKS power spectrum with \citet{S95} shape parameter.

\section{Halos and Peaks}\label{peak-halo}

In hierarchical cosmologies with Gaussian density perturbations, DM clustering is fully determined by the linear power spectrum. Thus, by filtering the density field at any arbitrary small time $\ti$ with varying graining scales so as to uncover the collapsing patches of different masses, it should be possible to reconstruct the growth, abundance, and inner properties of halos at any time $t>\ti$. But is there really any smoothing window able to do that?

Given the isotropy of the Universe, the window must be spherical. On the other hand, as collapsing patches are finite, the window must be of compact support. Lastly, even though we do not understand why (see Sec.~\ref{mergers} for the explanation), simulations with finite resolution converge, so the window must also be of compact support in Fourier space. The Gaussian is the only window satisfying these necessary conditions, and as shown next it works, indeed.

\subsection{Ellipsoidal vs. Spherical Collapse}

In top-hat spherical collapse, all peaks with a fixed positive density contrast $\delta\F$ at any scale $R\F$ in the density field at $\ti$ collapse at the same time $t$, and give rise to halos with mass $M$ satisfying (e.g. \citealt{P80})
\beqa
\delta\F(t,\ti)=\delta\cc\F(t)\frac{D(\ti)}{D(t)}~~~~~~~~~~~~~~~~~~~~~~~~~~~~~~~~~~
\label{deltatF}\\
R\F(M,\ti)=\left[\frac{3M}{4\pi \rho\cc(\ti)}\right]^{1/3}\,.~~~~~~~~~~~~~~~~~~~~~~~~~~~
\label{rmF}
\eeqa
In equations (\ref{deltatF}) and (\ref{rmF}) $\rho\cc(t)$ is the cosmic mean density at $t$, $\delta\cc\F(t)$ and $D(t)$ are the critical linearly extrapolated density contrast for spherical collapse at $t$, and the linear growth factor (equal to 1.686 and to the scale factor $a(t)$, respectively, in the Einstein-de Sitter cosmology). The relation (\ref{rmF}) is equivalent to the following one in terms of the 0th-order top-hat spectral moment $\sigma_0\F$,
\beq
\sigma_0\F(M,\ti)=\sigma_0\F(M,t)\frac{D(\ti)}{D(t)}\,,
\label{sigmaF}
\eeq
which ensures that the seed of one halo at different $\ti$ is the same evolving perturbation with height $\nu\F(M,t)\equiv \delta\F(t,\ti)/\sigma_0\F(M,\ti)=\delta\F\cc(t)/\sigma_0\F(M,t)$. 

However, in Gaussian density fields, peaks are triaxial (for whatever filter), so the patches they encompass undergo ellipsoidal collapse, which invalidates the previous relations. Indeed, the time of collapse (along all three axes) $t\cc$ depends in this case not only on the mass and size, but also on the triaxial shape and central concentration of the patch, i.e. on the density contrast $\delta$, scale $R$, ellipticity $e$, prolateness $p$, and curvature $x$ (defined in eq.~[\ref{0th}]) of the associated peak: $t\cc=t\cc(R,\delta,e,p,x)$. It is thus not unsurprising that peaks with identical $\delta\F(t)$ but different $R\F(M)$ collapse at different times, and that the masses of spherical patches with $R\F(M)$ differ from the masses $M$ of halos resulting from their collapse (D i).

Nonetheless, as the $e$ and $p$ probability distributions functions (PDFs) of Gaussian-filtered peaks with $\delta$ and $x$ at scale $R$ are very sharply peaked at their maximum values like the $x$ PDF itself (BBKS), we can take them with fixed values at the respective maxima, $e\maxi$, $p\maxi$, $x\maxi$. Thus, provided that there is some relation between the mass $M$ of halos at the cosmic time $t$ (or the corresponding patches at $\ti$) and the scale $R$ of their associated peaks as provided by any given halo mass definition (see below), all ellipsoidal patches with different masses $M$ identified by triaxial peaks with a positive $\delta$ at the scale $R(M,t)$ will collapse essentially at the same cosmic time $t=\ti+t\cc[R(M,t),\delta,e\maxi,p\maxi,x\maxi]$, with $t$ a monotonous decreasing function of $\delta$ as in top-hat spherical collapse (D v). Certainly, the small scatter $\Delta$ in the $e$, $p$, and $x$ PDFs around the respective maxima will translate into a scatter in the time of collapse,
\beq
\Delta t\cc=\Delta t=\frac{\partial t\cc}{\partial e}\Delta e+\frac{\partial t\cc}{\partial p}\Delta p+\frac{\partial t\cc}{\partial x}\Delta x,
\eeq
which will propagate into the MF of halos at $t$, 
\beq
\Delta N(M,t)=\frac{\partial N}{\partial t}\Delta t,
\eeq
as well as on the radius $r$ of the ancestor at $t$ of the halo with final mass $M$, implying (in halos growing inside-out; see below), a scatter in the value at $r$ of any halo profile,   
\beq
\Delta \xi(r)=\frac{\der \xi}{\der r}\frac{\partial r(t,M)}{\partial t}\Delta t.
\eeq
To calculate those scatters we need the partial derivatives of $t\cc$ with respect to $e$, $p$, and $x)$, which can be estimated through numerical experiments. But the purpose of the present Paper is not to derive such scatters, but the mean halo properties, and this can be done analytically.

\subsection{Halo-Peak Correspondence and Halo Mass Definition}
\label{correspondence}

Thus, for any given halo mass definition, Gaussian ellipsoidal collapse leads to a one-to-one correspondence between halos with $M$ at $t$ and peaks in the density field at $\ti$ filtered on scale $R$ similar to that found in top-hat spherical collapse. Moreover, to guarantee the arbitrariness in $\ti$ the functions $\delta(t)$ and $R(M,t)$ defining that correspondence must depend on $\ti$ just as in top-hat spherical collapse,
\beqa
\delta(t,\ti)=r_\delta(t)\,\delta\F(t,\ti)\label{deltat}~~~~~~~~~~~~~~~~~~~~~~~~~~~~~~~~\\
R(M,t,\ti)=r_{\rm R}(M,t)\,R\F(M,\ti)\,,~~~~~~~~~~~~~~~~~\label{rm}
\eeqa
or, using the 0th-order Gaussian spectral moment instead of the scale $R$, 
\beq
\sigma_0(M,t,\ti)=r_\sigma(M,t)\,\sigma_0\F(M,\ti)\,.
\label{sig}
\eeq
Hereafter $\sigma_{\rm j}$ stands for the jth Gaussian spectral moment. Note that the halo-peak correspondence does not depend on the small scale mass distribution within each patch determining whether the collapse is monolithic or lumpy (i.e. Q ii).

Specifically, in all cosmologies, the consistency conditions that all the DM in the Universe must be locked inside halos of different masses and that the mass $M$ of a halo must be equal to the volume-integral of its density profile allow one to find \citep{Jea14a} the functions $r_\delta(t)$ and $r_{\rm R}(M,t)$ or $r_\sigma(M,t)$ setting the correspondence associated with every specific halo mass definition (D iii).\footnote{The usual ones are those dubbed SO($\Delta$) and FoF($b$). In the former, the mass defines a spherical overdensity $\Delta$ with respect to the mean cosmic density $\rho\cc(t)$ (for instance, the virial mass $M\vir$ corresponding to the virial overdensity $\Delta\vir(t)$; \citealt{BN98}) or to the critical cosmic density $\rho\crit(t)$ (for instance, $M_{200}$ corresponding to $\Delta=200$). In FoF($b$), the mass encompasses all DM particles that coalesce through the Friends-of-Friends algorithm with linking length $b$ in the density field at $t$.} In all cases of interest, these functions are well-fitted by the analytic expressions,
\beqa 
r_\delta(t)=\frac{a^d(t)}{D(t)}~~~~~~~~~~~~~~~~~~~~~~~~~~~~~~~~~~~~~~~~~~~
\label{cc}\\
r_\sigma(M,t)=1+S(t)\left\{\frac{\delta(t,\ti)}{\sigma_0\F[R\F(M),\ti]}\right\}\,,~~~~~~~
\label{sigma}
\eeqa
with $S(t)$ defined as
\beq
S(t)=s_0+s_1a(t)+\log
\left[\frac{a^{s_2}(t)}{1+a(t)/A}\right]\,,
\label{St}
\eeq
with the values of coefficients $d$, $s_0$, $s_1$, $s_2$ and $A$ for some relevant cases, quoted in Table 1. (See App.~\ref{approximation} for an approximate expression of $r_{\rm R}(M,t)$.)  

\begin{table}
\caption{Coefficients in the halo-peak relations.}
\begin{center}
\begin{tabular}{ccccccc}
\hline \hline 
Cosmology & Mass & $d$ & $s_0$ & $s_1$ & $s_2$ & $A$ \\ 
\hline
\multirow{2}{*}{Planck14$^a$} & $M\vir$ & 0.928 & 0.0226 & 0.0610 & 0.0156 & 11.7 \\ 
& $M_{200}$ &  0.928 & 0.0341 & 0.0684 & 0.0239 & 6.87\\ 
\multirow{2}{*}{WMAP7$^b$} & $M\vir$ & 1.06 & 0.0422 & 0.0375 & 0.0318 & 25.7\\ 
   &$M_{200}$ & 1.06 & 0.0148 & 0.0630 & 0.0132 & 12.4\\ 
\hline
\end{tabular}
\end{center}
$^{a}$ \citet{P14}\\
$^{b}$ \citet{Koea11}
\label{T1}
\end{table}

Strictly speaking, the previous halo-peak correspondence does not take into account that, due to cloud-in-cloud configurations, some nested peaks do not accomplish their collapse at $t$ because, before that, they are captured by the more massive halo corresponding to the host peak and become subhalos. Thus, the previous halo-peak correspondence must be corrected for failed halos and the corresponding nested peaks in order to have a strict one-to-one correspondence between real virialized halos and non-nested peak. In turn, similar one-to-one correspondences are foreseen between subhalos and nested peaks at different levels (see below).

\subsection{Halo Growth and Peak Trajectories}\label{growth} 

In hierarchical cosmologies, low-mass halos are much more abundant than high-mass ones, so minor mergers are extremely frequent, and give rise to a smooth mass increase of the massive partner called accretion. Accretion does not alter the quasi-equilibrium state of the accreting object; there is just a gentle virialization of the accreted matter through shell-crossing. In turn, accreted halos survive as subhalos. In contrast, major mergers are rare events causing the destruction of all partners (usually two) and the virialization ex novo of the whole system through violent relaxation. This relaxation takes a few crossing times, so a substantial fraction of all halos at any time $t$ (usually the most massive ones, with longer crossing times and formed more recently; e.g. \citealt{RGS01}) are not fully virialized. It is thus unsurprising that the seeds at $\ti$ of $\sim 15-20$\% of them still have multiple nodes \citep{Pea} (D ii). Thus, when calculating the halo MF at $t$, we will include {\it all} collapsed halos at that time, regardless of whether they are fully virialized or not, but when calculating their typical (steady) properties we will consider only virialized objects.

Taking into account the relation
\beq 
\frac{\partial\delta({\bf r},R)}{\partial R}= R \nabla^2\delta({\bf r},R)\equiv -x({\bf r},R)\sigma_2(R)R\,
\label{0th}
\eeq
satisfied by a Gaussian window, the Taylor expansion of $\delta({\bf r},R)$ at scale $R+\Delta R$ ($\Delta R \ll R$) of a peak implies there can only be one peak within a distance $\Delta R$ from another at scale $R$ \citep{MSS95}. We can thus readily identify the two peaks tracing any individual accreting halo. All peaks with varying $R$ tracing the same accreting halo describe a continuous $\delta(R)$ trajectory in the $\delta$--$R$ plane. Note that the peaks along any of those trajectories are not anchored to a fixed point; their position ${\bf r}$ slightly sloshes around. However, the series of (non-concentric) patches of different masses they encompass as $\delta$ diminishes are embedded within each other because the separation of contiguous peaks at $R$ and $R+\Delta R$ is at most $\Delta R$.

The continuous peak trajectory $\delta(R)$ traced by an accreting halo satisfies the differential equation
\beq
\frac{\der \delta}{\der R}=-x(\delta,R)\,\sigma_2(R) R\,,
\label{dmd}
\eeq
Similarly, the peak trajectory traced by a halo accreting at the {\it mean} instantaneous rate satisfies the same equation but with the curvature $x(\delta,R)$ replaced by the inverse of the mean inverse curvature of peaks with $\delta$ at $R$ \citep{MSS95} or, given the very sharply peaked $x$-distribution, simply the mean peak curvature $\hat x (R,\delta)$, calculated in BBKS.

For moderately high peaks as it corresponds to halos of galactic scales, $\hat x (R,\delta)$ is well-approximated by $\gamma \nu$, where $\gamma$ is $\sigma_1^2/(\sigma_0\sigma_2)$ (BBKS), and equation (\ref{dmd}) leads to
\beq
\frac{\der \ln\delta}{\der \ln R}\approx -\left[\frac{\sigma_1(R)}{\sigma_0(R)}\right]^2\,R^2\,.
\label{dmd2}
\eeq
Approximating the power spectrum by a power-law, $P(k)\approx Ck^n$, we have $\der \ln \delta/\der \ln R\approx m$ and $\delta(R)\propto R^m$, with $m=-[(n+3)/2]^{3/2}$, showing that the form of those peak trajectories is nearly universal in all realistic cosmologies. This result is what causes (App.~\ref{approximation}) all scaled halo profiles to be nearly universal (Q i). Another interesting consequence is that, in the limit of vanishing $R$, peak trajectories converge to a finite value.

Note that, contrarily to what happens in the excursion set (ES) formalism \citep{BCEK}, where the $\delta(R)$ trajectories describe random walks, the $\delta(R)$ trajectories in CUSP are continous monotonously decreasing functions of $R$ (eq.~[\ref{dmd}]). This is well-understood: ES follows the variation in the mass of patches around {\it fixed points} when the scale $R$ of the smoothing {\it $k$-sharp} window increases, whereas CUSP follows the variation in the mass of patches around {\it moving peaks} when the scale $R$ of the smoothing {\it Gaussian} window increases. In the former case the density contrast at the point may increase or decrease, while in the latter their curvature of the peak automatically determines how much $\delta$ decreases. (The curvature of the peak is indeed connected with the accretion rate of the corresponding halo; see eq.~[\ref{dmd}].) Far from being a problem, the monotonous decrease of $\delta(R)$ with increasing $R$ better reflects the monotonous increase of halo masses with increasing cosmic time. This difference between the two formalisms is what allows CUSP to monitor the growth of individual halos, while ES can only monitor the growth of halos in a statistical way (see also Sec.~\ref{mergers}). 

But, when an accreting halo suffers a major merger, its
associated peak trajectory is interrupted. No peak at $R+\Delta R$ can be identified to the peak at $R$ (in fact the two merging peaks become saddle points at the new scale; \citealt{Ciea20}) and a `new' peak with the same $\delta$ on a scale $R$ substantially larger than those of the progenitors appears (it cannot be identified to any peak at the previous scale $R-\Delta R$), which traces the newly formed halo. The rates at which peak trajectories disappear or appear due to major mergers can be used to calculate the halo destruction and formation rates (via these events) as well as their formation and destruction time PDFs \citep{Mea98}. The fact that, due to major mergers, the number density of (non-nested) peak trajectories in the $\delta$--$R$ plane progressively declines with increasing $R$ is
precisely what motivated this formalism to be dubbed {\it ConflUent System of Peak trajectories}.

\section{(Sub)Halo Mass Functions and Peak Abundances}\label{MF}

When a halo is accreted by another halo or merges with it, it becomes a subhalo or is destroyed, respectively. But its own first-level subhalos are never destroyed. Either they become second-level subhalos in the former case or they are passed as first-level subhalos in the new halo in the latter case. And so on for higher-level subhalos. 

Given the one-to-one halo-peak correspondence, this halo and subhalo behaviour automatically leads to a parallel behaviour of the corresponding peaks. 
The result is a complex nesting network of peaks with any fixed density contrast at different scales \citep{AJ90}. But this network is not a problem: it simply tells us that to count halos at any $t$ we must count non-nested peaks with the corresponding $\delta$ and to count subhalos of any desired level we must count nested peaks of the same level (D ii). 

\subsection{Halo MF and Abundance of Non-Nested Peaks}\label{numdens}

The number density of peaks with $\delta$ at scales between $R$ and $R+ \der R$ is the number density of peaks at scale $R$ with density contrast $\tilde\delta$ greater than $\delta$ that cross such a density contrast when the scale is increased to $R + \der R$ or, equivalently, with $\delta$ satisfying the condition
\beq
\delta < \tilde\delta \le \delta+\sigma_2(R)\,\tilde x\,R\,\der R\,.
\label{cond}
\eeq 
Therefore, it is equal to the integral of the density of peaks with height $\tilde \nu=\tilde \delta/\sigma_0(R)$ and curvature $\tilde x$ per infinitesimal $\tilde \nu$ and $\der \tilde x$, calculated by BBKS, over $\tilde \delta$ in the range delimited by the inequality (\ref{cond}) and over $\tilde x$ in the range of all possible values. The result is \citep{MSS95}
\beq
\!\!\!N\pk(R,\delta)\,\der R
=\frac{\hat x(R,\delta)}{(2\pi)^2R_\star^3}\,
\exp\left(-\frac{\nu^2}{2}\right)\,
\frac{\sigma_2(R)}{\sigma_0(R)}
\,R\,\der R\,,
\label{npeak}
\eeq
where $\nu=\delta/\sigma_0(R)$ is the peak height, and $R_\star$ is defined as $\sqrt{3}\,\sigma_1/\sigma_2$. As mentioned, the mean peak curvature, $\hat x(R,\delta)$, is separable in a first approximation, so is also $N\pk(R,\delta)$. 

But the number density (\ref{npeak}) refers to {\it all} peaks with $\delta$ between $R$ and $R+\der R$, while virialized halos correspond to non-nested peaks only. The number density of {\it non-nested} peaks with $\delta$ between $R$ and $R + \der R$, $N(R,\delta)\der R$ is the solution of the Volterra integral equation
\beqa
\!\!\!\!\!\!\!\!N(R,\delta)=N\pk(R,\delta)\nonumber~~~~~~~~~~~~~~~~~~~~~~~~~~~~~~~~~~~~~~~~~~~\\
\!\!\!-\frac{1}{\rho\cc(\ti)}\int_R^\infty dR' M(R')\,N(R',\delta)\,N\pk\nest(R,\delta|R',\delta),\;\;\;
\label{nnp}
\eeqa
where  $N\pk\nest(R,\delta|R',\delta)$ is the conditional density of peaks with $\delta$ at $R$ subject to being nested in a peak with $\delta'$ at $R'$ given below (eqs.~[\ref{mf16}]--[\ref{peaks2}]). The comoving density (\ref{nnp}) of peaks with $\delta$ at $R$ at $\ti$ is thus equal to the comoving density of halos with $M$ at $t$ or MF, $N(M,t)$, but for the change of variable from $R$ to $M$ (eq.~[\ref{rm}]). 

\begin{figure}
\centerline{\includegraphics[scale=0.46]{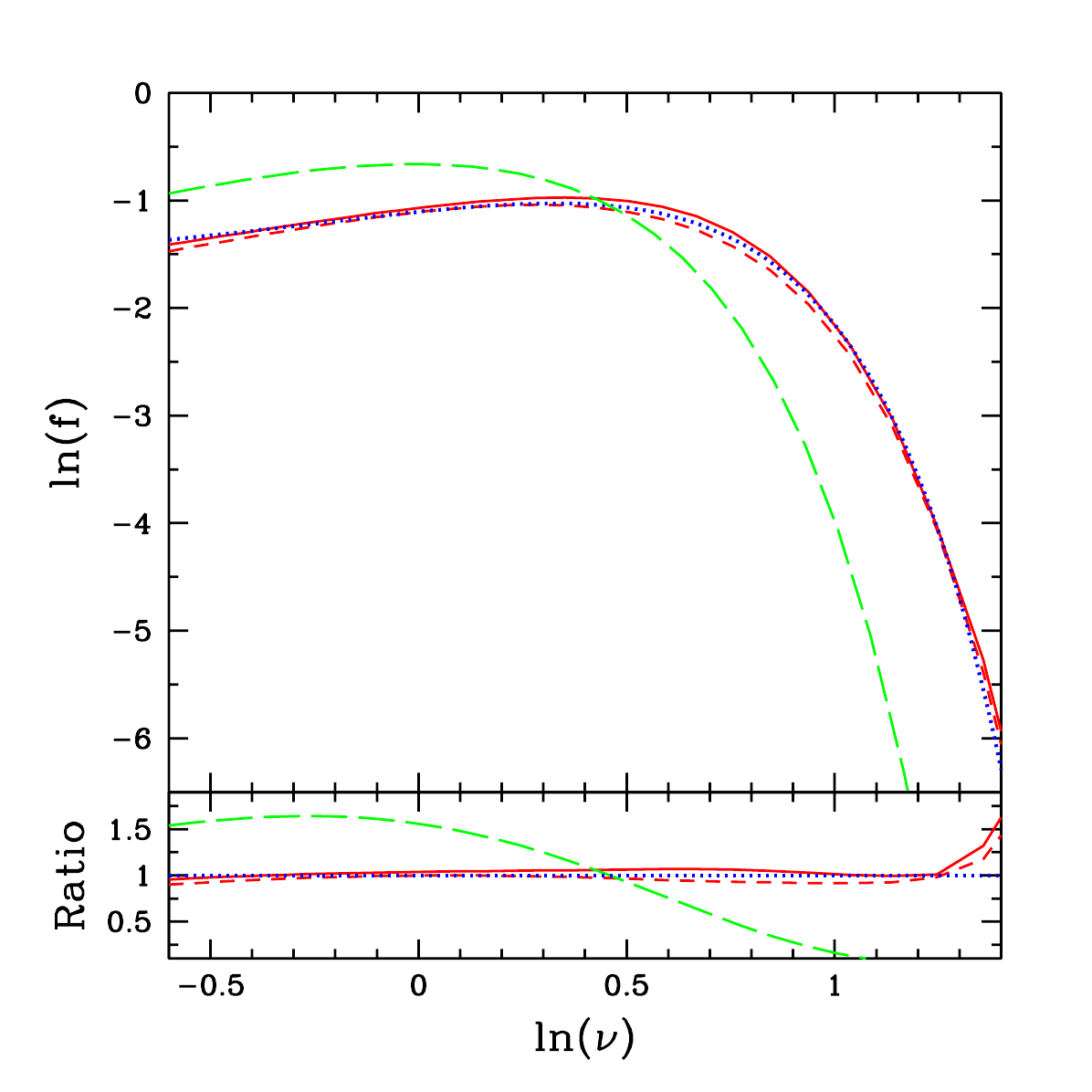}}
\caption{Multiplicity function of current halos predicted by CUSP with no free parameter (red lines), compared to the \citet{Wea06} analytic fit to the nearly universal multiplicity function of simulated halos (blue dotted line) for FoF(b=2) halo masses. The solid line is the solution given here \citep{Mea98} and the dashed line is a practical approximate solution \citet{Jea14b}. For comparison we also plot the Press \& Schechter (1974) MF (green long-dashed line).}
\label{f1}
\end{figure}

The new change of variable from $M$ to $\sigma_0$ leads to the halo multiplicity function at $t$, defined as
\beq 
f(\sigma_0,t)=\frac{M}{\rho\cc(t)}\frac{\partial N[M(\sigma_0),t]}{\partial \ln
  [(\sigma_0)^{-1}]}\,.
\label{f}
\eeq
And, using the variable $\nu=\delta\cc(t)/\sigma_0[R(M,t)]$ instead of $\sigma_0$, we arrive at 
\beq 
f(\nu,t)=\frac{M(\nu,t)}{\rho\cc(t)}\,
N(\nu,t)\,,
\label{CUSPmf}
\eeq
(see Fig.~\ref{f1}), with the Gaussian height $\nu$ approximately equal to $[a(t)/D(t)]$ times the top-hat one, $\nu\F$ (App.~\ref{approximation}). 

As shown in \citet{Jea14b}, the FoF mass with linking length $b=0.2$, FoF(0.2), is equivalent to the SO mass with overdensity $\Delta\vir$ relative to the mean cosmic density, SO($\Delta\vir$). This explains that the halo multiplicity function for FoF(0.2) is privileged (Q iii) in the sense that it has a roughly universal shape in top-hat filtering (\citealt{Jea14b}).

\subsection{Subhalo MF and Abundance of Nested Peaks}\label{abundances}

Following the same procedure above but from the conditional density of peaks with infinitesimal $\tilde \nu$ and $\tilde x$ at $R$ subject to lying at a distance $r$ (in units of the top-hat filtering radius associated with $R$) of a peak with $\nu'$ at $R'$, per infinitesimal $\tilde \nu$ and $\tilde x$ calculated by BBKS, we can also compute the conditional number density of peaks with $\delta$ at scales between $R$ and $R + \der R$ subject to being at a distance $r$ (same units) from such a peak
\beqa 
N\pk(R,\delta|R',\delta',r)\der R~~~~~~~~~~~~~~~~~~~~~~~~~~~~~~~~~~~~~~~~~~~\nonumber\\
=\!\!\frac{\hat x(R,\delta,R'\!,\delta'\!,r)}{(2\pi)^2\,R_{\star}^3\,e(r)}
      \!\exp\!\left\{\!-\frac{\left[\nu\! -\!
            \epsilon(r)\nu'(r)\right]^2}{2e^2(r)}\!\right\}\!\!{\frac{\sigma_2(R)}{\sigma_0(R)}}
      R\,\der R,~~~~
\label{mf16}
\eeqa 
where $\hat x(R,\delta,R',\delta',r)$ is the mean curvature of peaks with $\delta$ and $R$ at a distance $r$ from another peak with
$\delta'$ at $R'$, and $\nu'(r)$ stands for $\overline{\delta'(r)}/\sigma_0(R') g(r,R')$, where we have used the notation: $e(r)=\sqrt{1 - \epsilon^2(r)}$, $\epsilon(r)=\sigma_0^2(R_{\rm m})/[\sigma_0(R)\sigma_0(R')] g(r,R')$, $R_{\rm m}^2=R\,R'$, and $g(r,R')=\sqrt{1-[\Delta\delta'(r)]^2/\sigma_0^2(R')}$, where $\overline{\delta'(r)}$ and $\Delta\delta'(r)$ are the mean and rms density contrast, respectively, at $r$ from a peak (BBKS).

Thus, the conditional number density of peaks with $\delta$ at scales $R$ to $R + \der R$ subject to being nested in a peak with $\delta'$ at $R'$ is \citep{Mea98}
\beqa 
\!\!\!\!\!N\pk\nest(R,\delta|R',\delta')= C\int_0^1\!
\der r\, 3 r^2 N\pk(R,\delta|R',\delta',r),\,\,
\label{int}
\eeqa
where factor
\beqa
\!\!\!C\equiv\frac{4\pi S^3 N(R',\delta')}{3N\pk(R,\delta)}
\int_0^{S}\der r\,3 r^2\,N\pk(R,\delta|R',\delta',r),\,\,\,
\eeqa
with $S$ equal to the mean non-nested peak separation (same units) drawn from their mean density (eq.~[\ref{nnp}]), is to correct for the overcounting of host peaks as some of them are also nested.

But, if we are interested in the number of first-level subhalos in a halo (from now on, numbers are denoted by a calligraphic $N$ so as to distinguish them from number densities), the conditional number density (\ref{int}) is not enough. We need the conditional number of peaks with $\delta$ per infinitesimal scale around $R$ subject to being nested {it at first-level} in a non-nested peak with $\delta$ at $R'$. The result is  
\beq 
{\cal N}(R,\delta|R',\delta)=\frac{M(R')}{\rho\cc(\ti)}
N\nest(R,\delta|R',\delta)\,, 
\label{peaknum} 
\eeq 
where $N\nest(R,\delta|R',\delta)$ is the conditional number density of peaks with the same characteristics subject to being nested in non-nested peaks with identical $\delta$ at $R'$ (eq.~[\ref{int}]) {\it corrected for nesting at any intermediate scale $R''$ between $R$ and $R'$}, given by the solution of the Volterra equation
\beqa 
N\nest(R,\delta|R',\delta)=
N\pk\nest(R,\delta|R',\delta)~~~~~~~~~~~~~~~~~~~~~~~~~\nonumber\\
\!\!\!-\!\!\int_{R}^{R'}\!\!\!\der
R''N\pk\fnest\!(R,\delta|R'',\delta) N\pk\nest\!(R'',\delta|R',\delta)\,\frac{M(R'')}
{\rho\cc(\ti)}.~~~~~
\label{peaks2} 
\eeqa
Note that, in the integral on the right of equation (\ref{peaks2}), we have used the conditional number density of peaks with $\delta$ per infinitesimal scale around $R'$ subject to being {\it directly} nested within peaks at $R''$ (i.e. without being nested in any smaller scale peak), $N\pk\fnest(R',\delta|R'',\delta)$, calculated in \citet{Sea19a} to avoid overcorrection for intermediate nesting (peaks can be nested in more than one intermediate scale peak).

We remark that $\hat x (R,\delta,R',\delta,r)$, like $\hat x(R,\delta)$, is very nearly separable, in the subhalo mass range, in a function of $R$ and another function of the remaining arguments (\citealt{Jea14b}). This separability then propagates (see  \citealt{Sea19b} for details) to the conditional number of nested peaks, ${\cal N}(R,\delta|R',\delta)$ (eq.~[\ref{peaknum}]), which has the important consequences mentioned below. 

\begin{figure}
\centerline{\includegraphics[scale=.47,bb= 5 20 540 578]{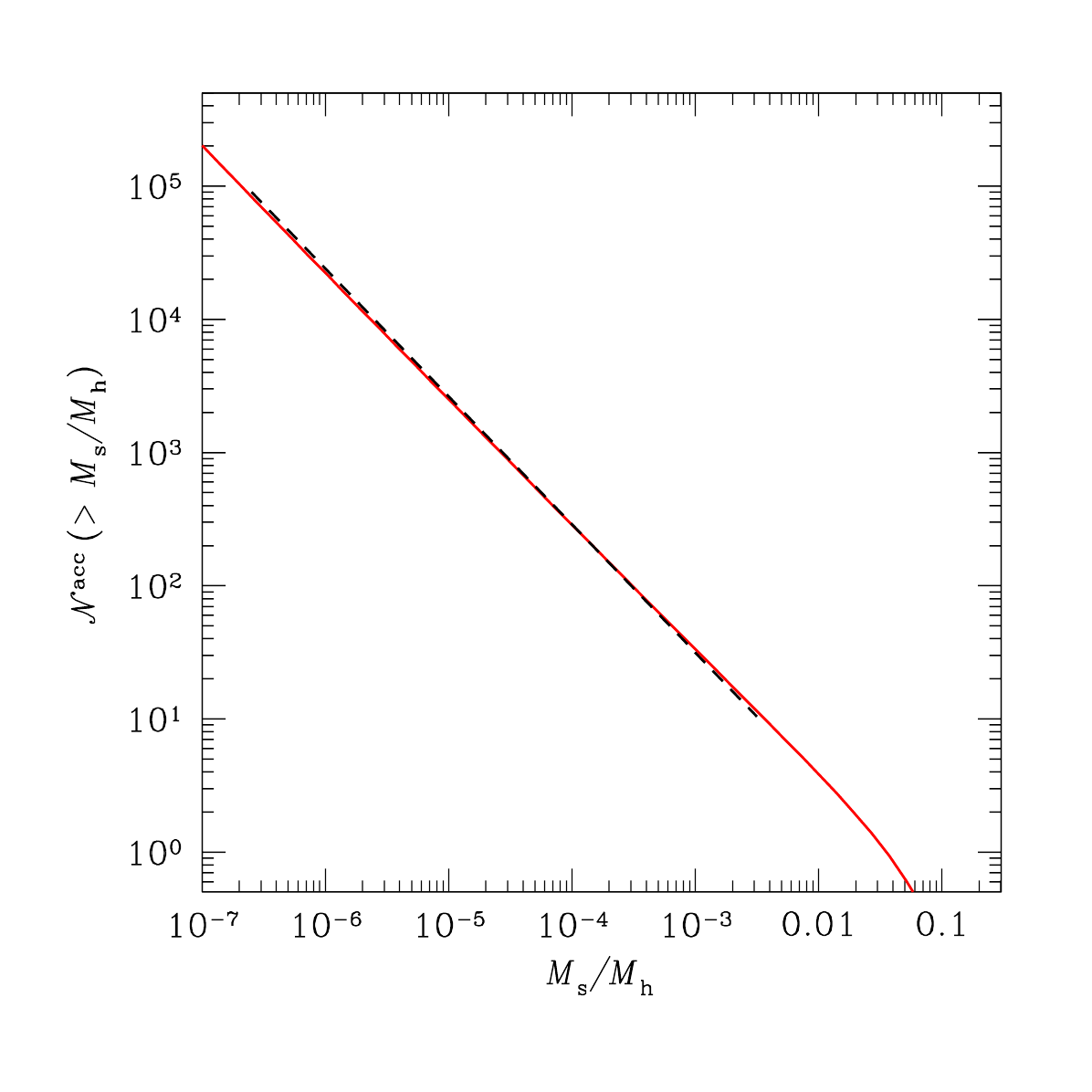}}
\caption{Cumulative MF of accreted subhalos as a function of subhalo mass predicted by CUSP with no free parameter (red line) for current halos with Milky Way mass ($M\h=2.2\times 10^{12}$ \modot), compared to the results of simulations by \citep{Han15} (dashed black line) found for the A halo in the Aquarius simulation \citep{Sea08}.}
\label{f2}
\end{figure}

Given the one-to-one correspondence between first-level subhalos and first-level nested peaks, the MF or number of accreted subhalos per infinitesimal mass in halos with $M\h$ at $t\h$, ${\cal N}\acc(\clM)$, coincides, but for the change of variable from scale $R$ to subhalo mass $\clM$, with the number of peaks per infinitesimal scale around $R(\clM)$ and $\delta(t\h)$ that are {\it directly} nested into a peak with the same density contrast at the larger scale $R(M_h)$. The corresponding cumulative subhalo MF ${\cal N}\acc(>\clM)$ (see Fig.~\ref{f2}) is essentially proportional to $\clM^{-1}$ (Q x), and very nearly universal when $\clM$ is scaled to $M\h$ (Q i). The derivation of the MF of subhalos tidally stripped by the halo potential well is postponed to Section \ref{stripping}.

\section{Protohalo Properties}\label{protohalo}

The typical halo properties, hereafter denoted by subscript h, arise from those of the corresponding collapsing patches or protohalos, hereafter denoted by subscript p. (Quantities with no subscript refer to both objects indistinctly.)
Specifically, since in linear regime the velocity and density fields are tightly related, protohalos are fully characterized by their spherically averaged density, ellipticity, and prolateness profiles, that is we need no kinematic information. To obtain those structural protohalo profiles we should deconvolve the height, ellipticity, and prolateness of the corresponding peaks. But this is not possible because the Gaussian window used to find those peaks yields the loss of information at large wavenumbers. Fortunately, as shown next, for {\it purely accreting halos} we can use the information on the peaks along the full $\delta(R)$ trajectory to achieve the desired deconvolutions. (The peak trajectories of ordinary halos suffering major mergers are interrupted at those events, so we have no access to that information.)

For this reason we concentrate, until Section \ref{mergers}, in halos evolving by pure accretion (PA), the only ones for which we can determine the properties of the corresponding protohalos. Then, monitoring the collapse and virialization of those protohalos, we can derive the properties of the final purely accreting halos. This procedure can be followed for individual objects (provided their $\delta(R)$ trajectory is known) or for ensembles of halos with the same mass and redshift so as to derive their typical properties. In the present study we concentrate in this latter application of the method. Specifically, we derive the {\it mean spherical averaged radial profiles and global quantities} of (proto)halos with $M\h$ at $t\h$.

\subsection{Density Profile}\label{density}

Taking the origin of the coordinate system on the peak at scale $R$, the density contrast $\delta$ at ${\bf r}\p=0$ is nothing but the convolution with the Gaussian window of that radius of the {\it intrinsic} density contrast field $\delta\p({\bf r}\p)$ in the protohalo. We thus have
\beq 
\delta(R)=\sqrt{\frac{2}{\pi}} \frac{1}{R^3} \int_0^{\infty}\der r\p\,
r\p^2\,\delta\p(r\p)\,{\rm exp}\left(-\frac{r\p^2}{2R^2}\right)\!,
\label{dp1}
\eeq 
after integrating over the polar angles, where $\delta\p(r\p)$ is the spherical average of $\delta({\bf r})$. Equation (\ref{dp1}) shows that we can infer the peak trajectory $\delta(R)$ traced by a purely accreting halo from the density profile of its protohalo. Conversely, given the peak trajectory $\delta(R)$ of a purely accreting halo, we can readily solve the Fredholm integral equation of first kind (\ref{dp1}) for $\delta\p(r\p)$ \citep{Sea12a}. This can be done for individual halos or for halos with $M\h$ at $t\h$ accreting at the mean instantaneous rate, leading to the {\it mean} spherically averaged density profile $\delta\p(r)$ of the corresponding protohalos. Note that, as in the limit of vanishing $R$, the peak trajectories converge to a finite value (null asymptotic slope), so do the unconvolved mean protohalo density profiles in the limit of vanishing $r\p$.

\subsection{Eccentricity Profiles}\label{protoshape}

Similarly, reorienting the Cartesian axes $j$ along the main axes of the triaxial peak with $\delta$ at $R$, the squared semiaxes of the peak, $A\jpk^2(\delta,R)$, equal to the second order spatial derivatives $\partial^2_j$ of the centered density contrast field $\delta\p({\bf r}\p)$ scaled to the Laplacian ($A\opk^2+A\tpk^2+A\zpk^2=1$) is given by \citep{Sea12b}
\beqa
A^2_j(\delta,R) \frac{x(\delta,R)}{\sigma_2(R)}=-\frac{1}{(2\pi R^2)^{3/2}}~~~~~~~~~~~~~~~~~~~~~~~~~\nonumber\\
\times \partial_j^2\bigg\{\!\!\int\! \der
  {\bf r}\p\, \delta\p({\bf r}\p)\exp\!\left[-\frac{({\bf r}\p\!-\!{\bf r'})^2}{2R^2}\right]\!\!\bigg\}_{\!{\bf r'}=0}.\label{lambda}
\eeqa
Writing $\delta\p({\bf r}\p)$ in terms of $\delta\p(r\p)$ and the protohalo axis profiles $a\jp(r\p)$ (App.~\ref{eccentricity}), and integrating over the polar angles, we arrive at
\beqa
A\jpk^2\!\left(\frac{\partial \delta}{\partial R},R\right)\,\frac{\partial \delta}{\partial R}
  + 3\,\delta(R) \!=\!\sqrt{\frac{2}{\pi}}\frac{3}{5R^5}\!\!~~~~~~~~~~~~~~~~~\nonumber\\
\!\!\!\!\!\!\!\!\times\!\!\int_0^\infty\!\! \der
r\p\, r\p^4\delta\p\!(r\p)\!\left[\frac{2a\po^2\!(r\p)}{a\jp^2\!(r\p)G\p(r\p)}\!+\!1\right]\!\exp\!\left(\!-\frac{r\p^2}{2R^2}\!\right)\!,
\label{inv}
\eeqa
where we have taken into account that the axes $A\jpk$ of peaks are are well-known functions of their curvature (BBKS), and introduced the function $G\p(r\p)$ defined in Appendix \ref{eccentricity}.

Equation (\ref{inv}) shows that the $A\jpk(R)$ trajectories of peaks can be obtained from the semiaxes profiles of their protohalos. Conversely, given a peak trajectory $\delta(R)$ determining the $A\jpk(R)$ trajectories, we can solve Fredholm integral equation of the first kind (\ref{inv}) for the quantities $a^2\po(r\p)/a^2\jp(r\p)$, in the same way as equation (\ref{dp1}). Then, from the quantities $a^2\po(r\p)/a^2\jp(r\p)$, we can infer the ellipticity and prolateness profiles, or equivalently, the primary and secondary eccentricity profiles, $\ep\p(r\p)$ and $\es\p(r\p)$, defined in Appendix \ref{eccentricity}. We can thus invert equations (\ref{inv}) from $\delta(R)$ solution of equation (\ref{dmd2}) (with the partial derivative $\partial \delta/\partial R$ equal to the total derivative) fixing the $A^2\jpk(\der \delta/\der R,R)$ trajectories, and determine the profiles $a^2\po(r\p)/[a^2\jp(r\p)G\p(r\p)]$ for the three orientations j, leading to the primary and secondary eccentricity profiles, $\ep\p(r\p)$ and $\es\p(r\p)$, for individual protohalos or for halos accreting at the mean instantaneous rate.

\section{Halo Properties}\label{accretion}

The properties of purely accreting halos follow from those of the corresponding protohalos thanks to the conservation in ellipsoidal collapse and virialization of a few quantities specified below. In fact, as the latter properties follow from the peak trajectories traced by such halos, the properties of purely accreting halos ultimately follow from the characteristics of those peak trajectories. 

The link between the properties of protohalos and halos is provided by the conservation of three quantities during ellipsoidal collapse and virialization: the mass inside the corresponding radii, the total energy inside those radii, and the volume delimited by them. The latter two quantities are not conserved in an absolute way (the system expands and contracts, and loses energy through shell-crossing during virialization), but relative to those of the spherically symmetrized system. 

\subsection{Density Profile}\label{halodensity}

The one-to-one correspondence between halos and peaks guarantees the conservation of the mass in the triaxial protohalo at $\ti$ to that of the triaxial halo at $t$, $M\h=M\p$. As shown in Appendix \ref{approximation}, this equality and the halo mass definition implicit in that correspondence fully determine the (roughly universal) mean spherically averaged halo density profile in purely accreting halos. However, to understand what is behind that formal derivation it is convenient to monitor in detail the ellipsoidal collapse and virialization of the system. For the reasons explained in Appendix \ref{sphericalApp}, this can be done assuming spherical symmetry.

In linear regime, homeoids expand radially by the factor $D(t)$ (hence, homothetically) with neither shell-crossing nor energy exchange between different regions. After reaching turnaround, shells collapse and bounce, which causes them to cross other shells. When a shell moving inwards crosses another one moving outwards some (potential) energy is transferred from the latter to the former due to the change in the inner mass they see. And when they again cross in the opposite direction the energy exchange is reversed. But the amount of energy exchanged depends on the radius of the crossing: near pericenter, they exchange more energy than near apocenter. This difference causes a net energy flux outwards, i.e. from shells collapsed earlier, with smaller apocenters, to shells collapsed later, with larger apocenters. This energy loss of shells causes their apocenter loci to shrink so shells cross increasingly nearer the center, and exchange less energy. On the other hand, the correlation between their phases is slowly lost. Thus, the Lagrangian energy outflow diminishes. In the end, the energy exchange ceases, and shell apocenters stop contracting. 

\begin{figure*}
\vspace{-0.4cm}
\hspace{12pt}
\includegraphics[scale=0.41]{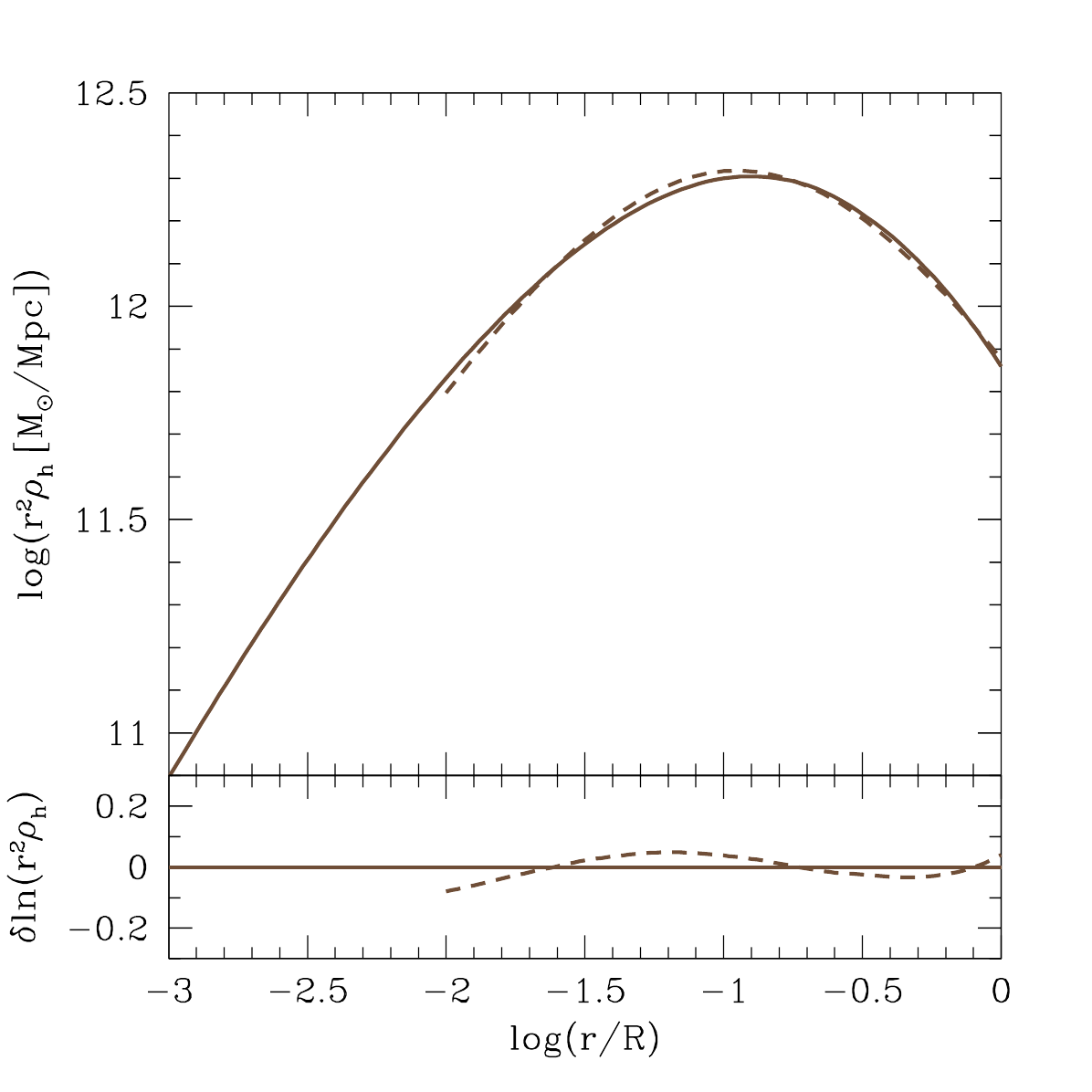}
\hspace{-8pt}
\includegraphics[scale=0.41]{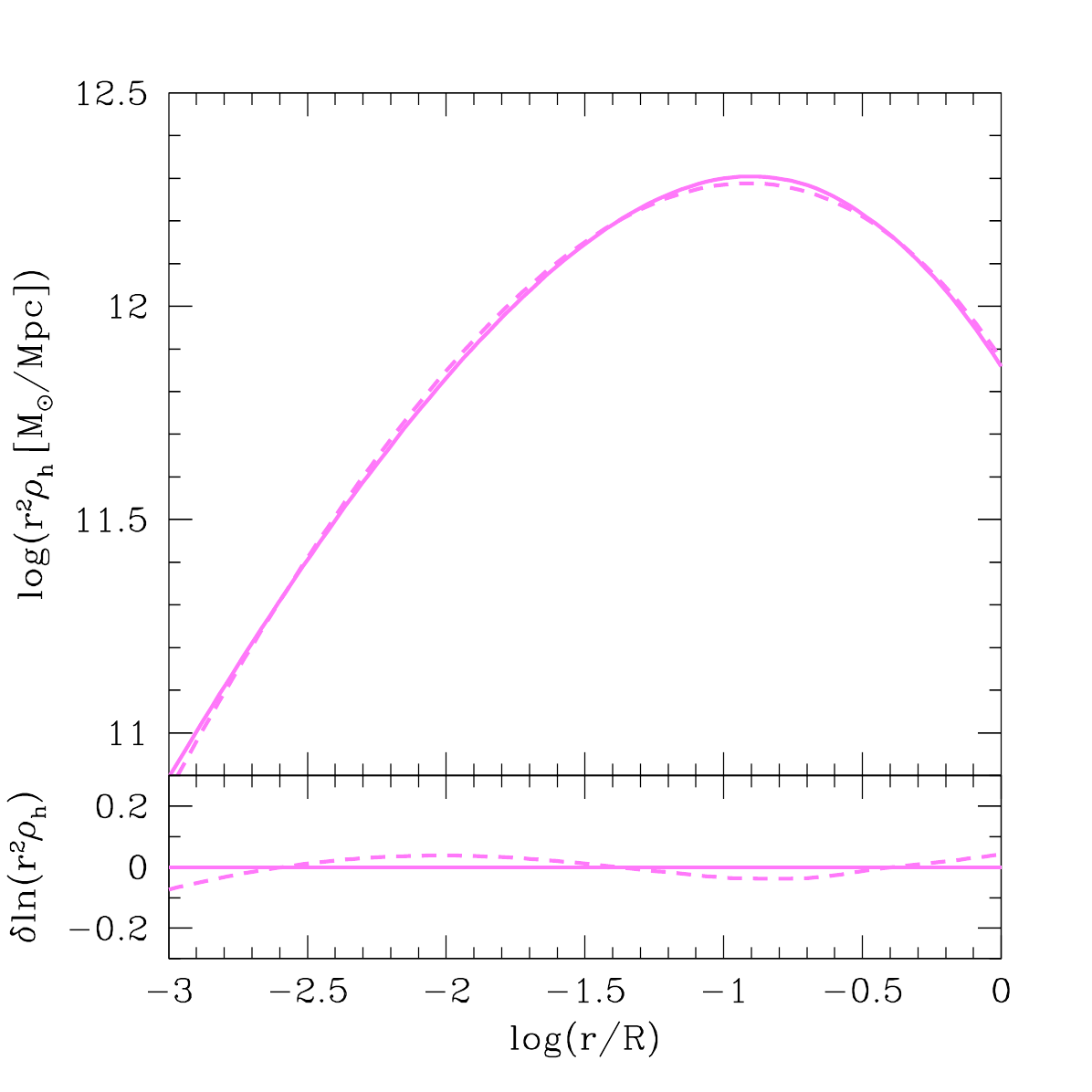}
 \caption{Mean spherically averaged density profile predicted by CUSP with no free parameter (solid lines) for current halos with $M\vir=10^{13}$ \modot, compared to usual analytic fitting expressions (dashed lines) for halos of that mass: the NFW \citep{NFW95} profile (left panel) and the Einasto (1965) profile (right panel). The residuals in the bottom show the same slight S-shape as found in the fits to the numerical profiles found in simulations.}
\label{f3}
\end{figure*}

Note that shell-crossing proceeds with no apocenter crossing. Indeed, if particles in two different accreting shells did coincide at apocenter (hence with null radial velocity), they will coincide along the whole trajectory, so they would not belong to different accreting shells. Therefore, the particle apocenter loci contract, due to the energy loss through shell-crossing, in an orderly manner until they stop. In other words, {\it accreting halos grow from the inside out} (D vi). This particular growth allows one to relate the radius $r\h$ in the final object in equilibrium to the protohalo properties within the corresponding $r\p$.

The virial radius of the accreting halo increases with increasing time as the object grows inside-out. We can thus {\it virtually} move one after the other the shells reaching turnaround without any crossing so as to match their apocenters in the final halo.\footnote{By `virtual' motion we mean a motion where the energy is conserved, although not the timing of the real motion.} The energy profile of the resulting `toy object' inside any radius $r\h$, $\widetilde E\h(r\h)$ will differ, of course, from that of the real halo $E\h(r\h)$ because the latter accounts for the energy loss through shell-crossing, while the toy object is built with no shell-crossing. In fact, $\widetilde E\h(r\h)$ equals the total energy of the system at turnaround or directly at the initial time, $\widetilde E\h(r\h)=E\p(r\p)$. The toy object is of course not in equilibrium, but this can be fixed. As $\widetilde E\h(r\h)-\widetilde W\h(r\h)$, where $\widetilde W\h(r\h)$ is the potential energy of a homogeneous sphere with $M\h(r\h)$ at $r\h$, must be positive,\footnote{The system at turnaround is less concentrated than the homogeneous toy object.} we can expand virtually every inner shell, avoiding shell-crossing, so as to end up with a uniform density equal to the mean density of the real halo inside $r\h$, and still have an excess of spherical kinetic energy inside that radius. This kinetic energy can then be redistributed over the sphere, exchanging the radial and tangential components of the spherical velocity variance so as to satisfy the spherical virial relation with null spherical radial velocity variance. We are then led to a steady homogeneous toy object with the same mass $M\h(r\h)$ and radius $r\h$ as the real halo, though with potential energy ${\widetilde W\h(M\h)}=-3/5\,GM\h^2/r\h(M\h)$, total energy $\widetilde E\h(M\h)=E\p(M\p)$, and null surface term. It thus satisfies the virial relation
\beq 
r\h(M\h)= -\frac{3}{10}\,\frac{GM\h^2}{E\p(M\p)}\,,
\label{vir0}
\eeq
where $E\p(M\p)=\widetilde E\h(M\h)$ is the non-conserved total energy of the protohalo ($\widetilde E\h(M\h)\ne E\h(M\h)$), which is given, in the parametric form, by
\beqa 
\!\!\!\!E\p(r\p)\!=\!4\pi\!\!\int_0^{r\p}\!\!\!\der r\, r^2 \rho\p(r)\!\left\{\!\!\frac{\left[H_{\rm i} r\!-\!v\p(r)\right]^2}{2}\!-\!\frac{GM\p(r)}{r}\!\!\right\}\;\;\;\; \label{E1}
\\
M\p(r\p)=4\pi\int_0^{r\p} \der r\, r^2\, \rho\p(r)\,,~~~~~~~~~~~~~~~~~~~~~~~~~~~~~~~
\label{M1}
\eeqa
with $\rho\p(r\p)=\rho\cc(\ti)[1+\delta\p(r\p)]$ calculated in Section \ref{protohalo}. $H_{\rm i}$ is the Hubble constant at $\ti$, and
\beq
v\p(r\p)=\frac{2G\left[M\p(r\p)-4\pi r\p^3\rho\cc(\ti)/3\right]}{3H(\ti) r\p^2}\,
\label{vp}
\eeq
is the peculiar velocity caused by the mass excess within $r\p$, where we have neglected the velocity dispersion of DM particles at $\ti$, and taken the cosmic virial factor $f(\Omega)\approx \Omega^{0.1}$ at $\ti$ equal to one. Note that equation (\ref{vir0}) explains why the virial radius of halos is recovered in spherical top-hat collapse assuming energy conservation (Q v), as if the system had not lost energy through shell crossing. This result thus justifies the definition of the halo virial radius \citealt{BN98}.

The inversion of $r\h(M\h)$ given by Equation (\ref{vir0}) 
leads to the mass profile $M\h(r\h)$, and by differentiating it at the spherically averaged density profile $\rho\h(r\h)$ (see Fig.~\ref{f3}). Conversely, given a purely accreting halo with mass profile $M\h(r\h)$, we can calculate the total energy of its seed, $E\p(M\p=M\h)$ (eq.~[\ref{vir0}]), and determine $\rho\p(r\p)$ from (eqs.~[\ref{E1}] and [\ref{M1}]). Thus there is in PA {\it a one-to-one correspondence between the density profile for halos and protohalos}. These previous results hold for the profiles of individual halo-protohalo pairs as well as for mean profiles. 

In addition, the inner asymptotic protohalo density profile, $\rho\p\propto r\p^\alpha$, implies $M\p(r\p)\propto r\p^{(3+\alpha)}$ and $E\p(r\p)\propto r\p^{(5+\alpha)}$. Equation (\ref{vir0}) then leads to $\rho\h(r\h)\propto r\h^\alpha$, and, since $\alpha$ is null (see Sec.~\ref{density}), we conclude that there is strictly no cusp in the mean spherically averaged density profile for CDM halos (Q iv). Nevertheless, the finite central value is approached slightly more slowly than in the Einasto profile.

\subsection{Eccentricity and Kinematic Profiles}\label{shape}

The kinematics and triaxial shape of halos are determined by the triaxial shape of protohalos. Since this relation is not included in the one-to-one halo-peak correspondence, there is no shortcut in this case; the only way to derive it is by taking into account the above mentioned conservation relations during ellipsoidal collapse and virialization. Moreover, for obvious reasons, we cannot assume spherical symmetry in this case. However, we will take into account that, to leading order in the the departure from spherical symmetry, hereafter simply the `asphericity', all physical properties of evolving triaxial systems coincide with those of their spherically symmetrized counterparts (see App.~\ref{sphericalApp}). 

To leading order in the asphericity, the volume of ellipsoids coincides at any time with that of the corresponding spheres in the spherically symmetrized system. That is, the ratio between the two volumes must be conserved, to leading order in the asphericity, over the ellipsoidal collapse and virialization of the system. We thus have
\beq
\frac{a\ho(r\h)\,a\htt(r\h)\,a\hz(r\h)}{a\po(r\p)\,a\ptt(r\p)\,a\pz(r\p)}=\frac{r\h^3}{r\p^3}\,,
\label{emass}
\eeq
where $a_1$, $a_2$, and $a_3$ are the semiaxes of the ellipsoids and $r$ is their equivalent radius (see the definition in App.~\ref{eccentricity}), equal to the radius of the corresponding sphere in the spherically symmetrized system.

Similarly, to leading order in the asphericity, the total energy within radius $r$ of the triaxial system, $E(r)$, coincides at any time with that ${\cal E}(r)$ in the spherically symmetrized system. Note that the energy lost by a sphere of initial radius $r\p$ through shell-crossing during virialization is indeed accounted for in ${\cal E}(r)$. What is not accounted for is the energy exchange between the sphere and the rest of the system in the triaxial system, so that energy loss is included in the residual $\delta{\cal E}(r)$. Consequently, the ratio between $E(r)$ and ${\cal E}(r)$ must be conserved to leading order in the asphericity. This conservation has two consequences. First, as $E(r)={\cal E}(r)+\delta{\cal E}(r)$, it implies
\beq
\frac{{\cal E}\h(r\h)}{{\cal E}\p(r\p)}=\frac{\delta
  {\cal E}\h(r\h)}{\delta {\cal E}\p (r\p)}\,,
\label{cale}
\eeq
with $\delta {\cal E}(r)$, given in Appendix \ref{sphericalApp}, dependent on the eccentricities of the halo. Second, as ${\cal E}(r)$ is the same in the spherically symmetrized system, the transfer from the radial to the tangential kinetic energy due to the non-radial motion produced in the non-linear evolution of the triaxial system must go in parallel to a departure of the potential energy so as not to alter the total spherical energy. In other words, the fractional velocity variance transferred from the radial to the tangential direction (equal to half the fractional 1-D tangential velocity variance generated) must be equal to half the typical fractional deviation of the potential,
\beq 
\frac{(\sigma^2\tang)\h(r\h)}{\sigma\h^2(r\h)}\!=\!\left\{\!\frac{\lav\delta\Phi\h\rav^2(r\h)}{\Phi^2\h(r\h)}\!\right\}^{\!1/2}.
\label{00th}
\eeq
Equation (\ref{00th}) relates the velocity anisotropy of halos with their eccentricity (Q vi).

\begin{figure}
\centerline{\includegraphics[scale=.43]{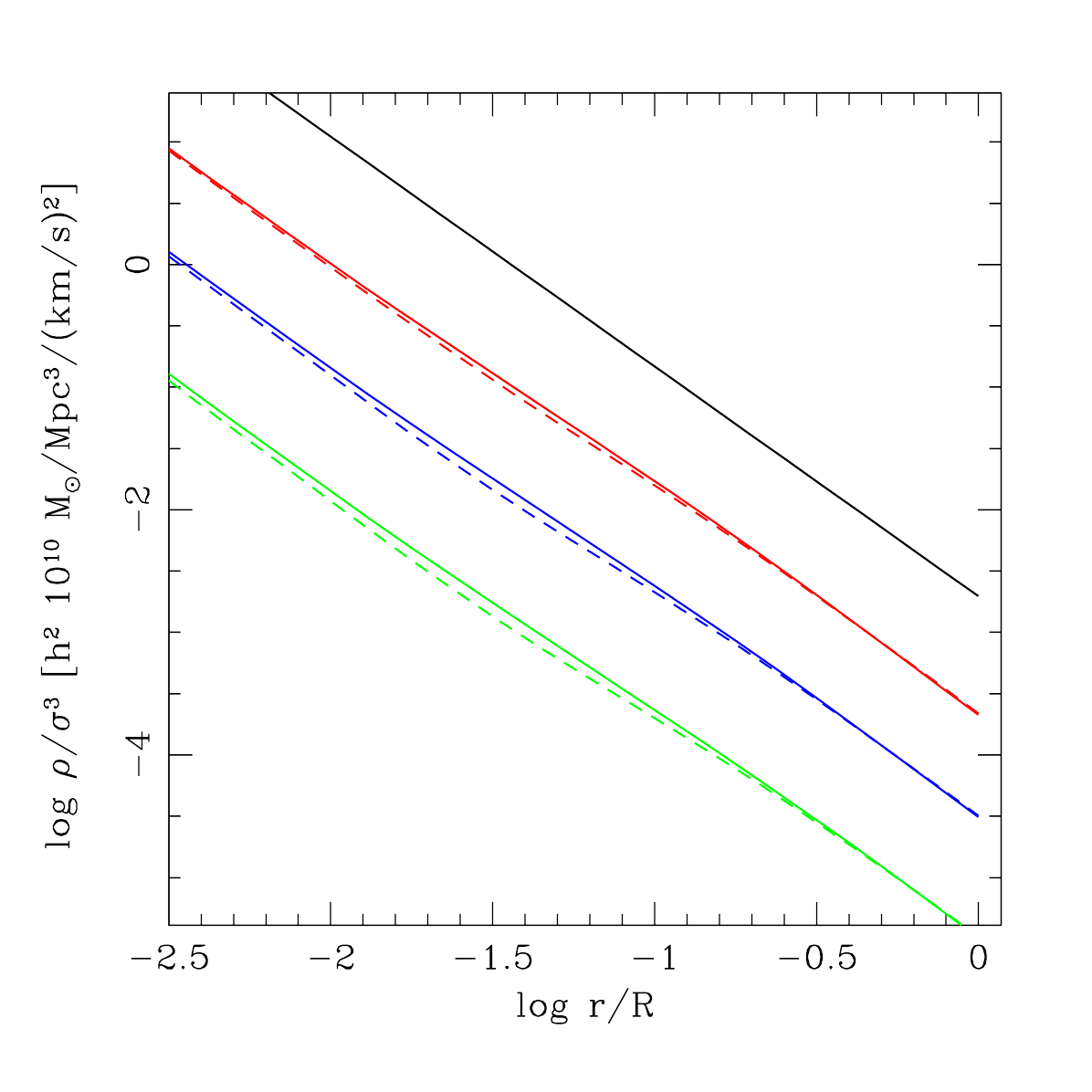}}
\caption{Mean pseudo phase-space density profiles predicted by CUSP with no free parameter (dashed lines) for current halos with the MW mass (red line), $10^{13}$ \modotb (blue line), and $10^{14}$ \modotb (green line). The solid colored lines are parallel to the black one with constant slope equal to $-1.875$ as found in simulations \citep{TN01}.}
\label{f5}
\end{figure}

Those two conservation relations determine the triaxial shape and kinematics of the halo (App.~\ref{eccentricity}). Indeed, the halo eccentricity profiles, $\ep\h(r)$ and $\es\h(r)$, arise from those of protohalos, $\ep\p(r)$ and $\es\p(r)$ (eqs.~[\ref{6th}] and [\ref{7th}]) through a relation (eq.~[\ref{S}]) that involves the halo velocity variance $\sigma\h^2(r)$. This relation is enough to determine the halo eccentricities near the center, where they coincide with those of protohalos with well-known sharply peaked PDFs (BBKS). But to solve the problem at any $r\h$ the closure relation (\ref{00th}) is required.

Writing the anisotropy $\beta\h(r\h)$ as a function of the scaled potential variance (eq.~[\ref{00th}]), the latter as a function of the halo eccentricities (App.~\ref{eccentricity}), and the eccentricities as functions of the halo velocity variance and the protohalo eccentricities (eqs.~[\ref{1st}] and [\ref{7th}]), the anisotropic equilibrium equation to leading order in the asphericity (App.~\ref{sphericalApp})
\beq
\frac{\der}{\der r\h}\!\left(\!\!\frac{\rho\h\, \sigma^2\h}{3\!-\!2\beta\h}\!\right)+\rho\h(r\h)\!
\!\left[\!\frac{2\beta\h}{3\!-\!2\beta\h}\frac{\sigma^2\h}{r\h}
\!+\!\frac{GM\h}{r\h^2}\right]\!(r\h)=0
\label{exJeq2}
\eeq
leads to a differential equation for $\sigma\h^2(r\h)$. Its solution for the usual boundary condition of null velocity at infinity leads to a pseudo phase-space density profile $\rho\h(r\h)/\sigma^3\h(r\h)$ of the same power-law form as in Bertschinger's (1985) spherical self-similar collapse model. This characteristic pseudo phase-space density profile found in simulations (see Fig.~\ref{f5}) is thus due to the (quasi-)self-similar collapse of a (quasi-) homogeneous mass distribution, and the phase-mixing produced by shell-crossing like in Bertschinger (1985) model (Q vii). The only difference, apart from the non-strict-similarity of the real collapse, is that Bertschinger assumes spherical symmetry, while CUSP takes into account the ellipsoidal collapse leading to a partial conversion of the radial velocities in linear regime to tangential velocities (eq.~[\ref{00th}]).

Once $\sigma\h(r\h)$ is known, we can compute the halo scaled density-potential covariance profile from that in the protohalo (eq.~[\ref{1st}]), and then the scaled density and potential variance profiles (App.~\ref{eccentricity}). The latter leads, through equation (\ref{00th}), to the velocity anisotropy profile $\beta\h(r\h)$, which adopts the universal form \citep{Hea06} found in simulations (Q viii).

These kinematic profiles lead to the eccentricity profiles of the halo from those of the protohalo (eqs~[\ref{1st}]-[\ref{S}]). In peaks with small and moderately large average curvature, the typical $e\p$ and $p\p$ values peak at $1.7$ and $1.3$, respectively, regardless of the values of $\delta$ and $R$ (BBKS), implying that $\ep\p$ and $\es\p$ are about $0.81$ and $0.64$, respectively. Since $U(r\h)$ is close to one (see Fig.~2 in \citealt{Sea12b}), the predicted mean $\ep\h(r\h)$ and $\es\h(r\h)$ profiles at small and moderately large radii are constant and close to $0.9$ and $0.8$, respectively, consistently with the results of simulations (Q ix). At large radii, the rms density fluctuation profile tends to a power-law of index $\sim -0.1$, the eccentricities are outwards decreasing, and the isopotential contours become more spherical than the isodensity contours \citep{Sea12b} as found in simulations (Q ix).

Once again, not only are the kinematic and eccentricity profiles of halos determined by the eccentricity profiles of the protohalo, but the latter can be obtained from the former (eqs.~[\ref{1st}]--[\ref{7th}]). And all the previous results hold for the profiles of individual halo-protohalo pairs as well as for their mean profiles.

\subsection{Subhalo Number Density Profiles}\label{truncation}

\subsubsection{Accreted Subhalos}\label{accreted}

According to the results of Section \ref{abundances} and the inside-out growth of halos in PA, the abundance of (first-level) subhalos per infinitesimal non-truncated mass around $\clM$ inside the radius $r\h$ in halos with current mass $M\h$ and virial radius $R\h$ at $t\h$, ${\cal N}\acc(<\! r\h,\clM)$, coincides, after the appropriate change of variables, with the mean number per infinitesimal scale of peaks with $\delta[t(r\h)]$ at $R(\clM)$ nested (at first-level) in the peaks with scale $R\{\delta[t(r\h)]\}$ found along the mean peak trajectory of those accreting halos, with mean (sperically averaged) mass profile equal to $M\h(r\h)$. We thus have,
\beqa 
{\cal N}\acc(r\h,\clM)=\frac{\der \wR}{\der\clM}\frac{\der {\cal N}[\wR,\delta|R(\delta),\delta]}{\der \delta}\frac{\der\delta}{\der R}\frac{\der R}{\der M\h}\frac{\der M\h}{\der r\h}\nonumber\\
= 4\pi\,r\h^2 \frac{\rho\h(r\h)}{M\h}\,{\cal N}\acc(\clM).~~~~~~~~~~~~~
\label{abun} 
\eeqa
Form now on a bar on a function of $r$ means its mean value inside $r$. 

The dDM mass fraction at $r\h$ in the halo with $M\h$ at $t\h$
directly accreted from the inter-halo medium is essentially equal to the dDM mass fraction $f\dc\acc(t)$ in that medium 
\beq
f\dc\acc(t)=1-\frac{1}{\rho\cc(t)}\int_{M\res}^\infty\!\der M\, M\,N(M,t)
\label{feq}
\eeq
($t>t\res$) at the time $t=t(r\h)$, being $N(M,t)$ the halo MF at $t$ (Sec.~\ref{MF}). Indeed, halo clustering starts at some finite time $t\res$ with a minimum halo mass $M\res$ equal to the free-streaming mass associated with the DM particle in the real Universe or the resolution mass in cosmological simulations. Consequently, all the DM that at $t\res$ should lie in halos with masses $M<M\res$ remains in the form of diffuse DM (dDM) that is progressively accreted onto halos (D viii) (see Fig.~\ref{f7}).  

As can be seen from equation (\ref{abun}), the above mentioned separability of ${\cal N}[\wR,\delta|R(\delta),\delta]$ in a function of $\wR$ and another function of the remaining arguments propagates (with an extra factor of $\wR$ arising from the $\delta$-derivative and the change of variable from $\wR$ to $\clM$) to the separability of the same kind of ${\cal N}\acc(r\h,\clM)$. 

The mean spherically averaged number density profile for accreted subhalos with $\clM$, $n\acc(r\h,\clM)$, defined as ${\cal N}\acc(r\h,\clM)/(4\pi r\h^2)$, scaled to the corresponding total mean number density, $\bar n\acc(R\h,\clM)=3{\cal N}\acc(\clM)/4\pi R\h^3$, is a function of $r$ only. Taking into account the relation (\ref{feq}), it can be written as
\beq
\frac{n\acc(r\h,\clM)}{\bar n\acc(R\h,\clM)}=\frac{\rho\h(r\h)}{\bar\rho\h(R\h)}
\label{dens}
\eeq
Equation (\ref{dens}) shows that the mean spherically averaged scaled number density profiles of accreted subalos for all masses $\clM$ overlap in one only profile, which follows the scaled density profile of the halo. The ultimate reason for this result is thus the above mentioned separability of the conditional number of nested peaks (Q xi).

\subsubsection{Stripped Subhalos}\label{stripping}

Soon after being accreted, subhalos begin to be tidally stripped (or even disrupted) by the host potential well so that they liberate other stripped subhalos and dDM previously locked inside them, and end up with smaller truncated masses, $\clM\tr$. Note that, as halos undergoing PA grow inside-out, once subhalos are accreted their orbits are kept unaltered, and since their initial conditions at accretion are independent of mass, the orbits are also. Certainly, subhalos suffer two-body interactions between themselves and with dDM. But these interactions do not affect the very numerous low-mass subhalos because they are already accounted for in shell-crossing (see the discussion in \citealt{Sea19a}). Only rare massive subhalos suffer an extra effect, the so-called dynamical friction. However, the typical properties of substructure are defined from large subhalo ensembles, so rare massive objects are excluded from the present treatment (D ix).   

The abundance ${\cal N}\str(r\h,\clM\tr)$ of stripped subhalos per infinitesimal radius and truncated subhalo mass around $r\h$ and $\clM\tr$ is then given by
\beqa {\cal N}\str(r\h,\clM\tr)\!=\!{\cal
  N}\tr(r\h,\clM\tr)\!+\!\bigg\lav\!\!\!\bigg\lav \!\int_{\clM}^{M\h(r\h)}\!\!\der M ~~~~~~~~~\nonumber\\  \!\!\!\!\!\times {\cal N}\acc(v,r\h,M)\!\!
\int_{R\tr(v,r\h,M)}^{R(r\h,M)}\!\!\! \der r{{\cal
    N}\str}_{[M,t(r\h)]}(r\!,\clM\tr)\!\!\bigg\rav\!\!\!\bigg\rav\,~~
\label{corr} 
\eeqa
where ${{\cal N}\str}_{[M,t(r\h)]}(r\!,\clM\tr)$ is the differential abundance of stripped subsubhalos with $\clM\tr$ at $r$ in a subhalo with mass $M$ when the host halo had mass $M(r\h)$. In equation (\ref{corr}), ${\cal N}\acc(v,r\h,M)$ is the abundance of accreted subhalos per infinitesimal mass, radius, and tangential velocity around $\clM$, $r\h$, and $v$, respectively, which factorizes in ${\cal N}\acc(r\h,M)$ (eq.~[\ref{dens}]) times the mass-independent tangential velocity distribution of the halo at $r\h$, and $R(r\h,M)$ and $R\tr(v,r\h,M)$ are the original and truncated radii of subhalos with original mass $M$ and tangential velocity $v$ at $r\h$, respectively. Double angular brackets denote average over the tangential velocity $v$ of subhalos at their apocentric radius $r\h$.

The first term on the right of equation (\ref{corr}) gives the abundance of subhalos accreted at $r\h$ with original non-truncated mass $\clM$ that give rise by direct stripping to subhalos with truncated mass $\clM\tr$, and the second term gives the number of subhalos with truncated mass $\clM\tr$ that are released at $r\h$ from the stripping of more massive subhalos.

Indeed, after subhalos are accreted (their apocenters are stabilized) at $r\h$, they begin to be repeatedly stripped and heated at each passage by the pericenter at $r_{\rm per}(v,r\h)$, causing them to reach a new equilibrium configuration at next apocenter. Taking into account the truncation undergone in every passage under the impulse approximation \citep{Gea94}, the subhalo ends up, after $p$ passages by pericenter (from $t(r\h)$ to $t\h$), with a truncated-to-original mass ratio given by \citep{Sea19b} 
\beq
\frac{\clM\tr(v,r\h,\clM)}{\clM}= m^p(r\h)\prod_{i=1}^{p} \left[\frac{Q_i(v,r\h)}{q(r\h)}\right]^3
\label{last}
\eeq
(Q xii), where the functions $m(r\h)$ and $q(r\h)$ can be calculated from the mass and radius those subhalos would have had they not suffered any stripping during virialization and $Q\ii(v,r\h)$ is the ratio of subhalo radii after and before truncation at the $i$-th pericentric passage. The ratio $Q\ii(v,r\h)$ is the solution of the equation
\beq
\frac{f[c_{i-1}(v,r\h) Q\ii(v,r\h)]}{f[c_{i-1}(v,r\h)]Q\ii^3(v,r\h)}=\frac{f[c(r\h)Q(v,r\h)]}{f[c(r\h)]Q^3(v,r\h)},
\label{mtr2}
\eeq
where $Q(v,r\h)$ stands for $r_{\rm per}(v,r\h)/r\h$ and $f(c)$ for $\ln(1+c)-{c}/(1+c)$ in the case of that the density profile of halos (or subhalos) is adjusted by the NFW \citep{NFW95} analytic profile (see \citealt{Jea14b} for the Einasto profile). In equation (\ref{mtr2}), $c(r\h)=r\h c(R\h)/R\h$ is the concentrations of the halo and  $c\ii(v,r\h)$ is the concentration of the subhalo when it reaches equilibrium at apocenter after the $i$-th pericentric passage is the solution of the recursive equation
\beq
\frac{h[c_{i+1}(v,r\h)]}{h[c_i(v,r\h)]}=u(v,r\h)\left[\frac{M_{i+1}(v,r\h,\clM)}{M_i(v,r\h,\clM)}\right]^{5/6}\!\!,
\label{rat2} 
\eeq
where $h(c)$ stands for $f(c)(1+c)/\{c^{3/2}[3/2-s^2(c)]^{1/2}\}$, being $s^2(c)$ the isotropic 3D velocity variance of objects with mass $M$ and radius $R$ scaled to $cf(c)GM/R$, and $u(v,r\h)$ is the heated-to-original subhalo energy ratio going together with tidal truncation, assumed of the form
\beq
u(v,r\h)={\rm max}\left\{1,K\left[\frac{M_{i+1}(v,r\h,\clM)}{M_i(v,r\h,\clM)}\right]^{\beta}\right\}.
\label{beta}
\eeq
For $K=0.77$ and $\beta=-0.5$ very good agreement is found with the results of numerical experiments (\citealt{Sea19b}). These two parameters are the only ones appearing in the whole theory. They are necessary because subhalo stripping and heating do not arise from the collapse and virialization of peaks, and hence, must be modeled apart.

According to the previous relations the truncated-to-original subhalo mass ratio (eq.~[\ref{last}]) turns out to be independent of the original mass of subhalos as a consequence of their similar concentration when they are accreted and the way they are truncated a pericenter (D xi)  (\citet{Sea19b}).

After some algebra, the solution of the integral equation (\ref{corr}) takes the compact form \citep{Sea19b}
\beq
{\cal N}\str(r\h,\clM\tr)=[1+f\st(r\h)]\,\mu(r\h)\,{\cal N}\acc(r\h,\clM\tr),
\label{M}
\eeq
where $\mu(r\h)$ is the $v$-averaged mass-independent truncated-to-original subhalo mass ratio profile and $f\st(r\h)$ is the proportion of stripped subhalos with $\clM\tr$ at $r\h$ previously locked as subsubhalos which have been released in the intra-halo medium when their subhalo hosts have been stripped. The quantity $f\st$, which can be found by solving a Fredholm equation given in \citet{Sea19b}, never exceeds 6 \%, so it can be neglected in a first approximation. 

Equation (\ref{M}) implies that the mean spherically averaged scaled number density profiles for stripped subhalos with $\clM\tr$, $n\str(r\h,\clM\tr)$ defined as usual in terms of the corresponding abundance ${\cal N}\str(r\h,\clM\tr)$ takes the form
\beq
\frac{n\str(r\h,\clM)}{\bar n\str(R\h,\clM)}=\frac{1-f\dc(r\h)}{1-\bar f\dc(R\h)} \frac{\rho\h(r\h)}{\bar\rho\h(R\h)},
\label{dens2}
\eeq
with $1-f\dc(r\h)$ defined as $[1-f\dc\str(r\h)]/[1-f\dc\acc(r\h)]$, where $f\dc\str(r\h)$ and $f\dc\acc(r\h)$ are the total dDM mass fraction and the dDM mass fraction of accreted subhalos, respectively, at the radius $r\h$ of the halo with $M\h$ at $t\h$ (D viii). The fraction $f\dc\str(r\h)$ is given by
\beq
f\dc\str(r\h)\!=\!1-\frac{1}{\rho(r\h)}\int_{M\res}^{M\h}\!\! \der \clM\tr\,\clM\tr\,\frac{{\cal N}\str(r\h,\clM\tr)}{4\pi r\h^2},
\label{rhod}
\eeq
and a similar relation holds for $f\dc\acc(r\h)$ in terms of ${\cal N}\acc(r\h,\clM)$.

\begin{figure}
\centerline{\includegraphics[scale=.45,bb= 18 50 560 586]{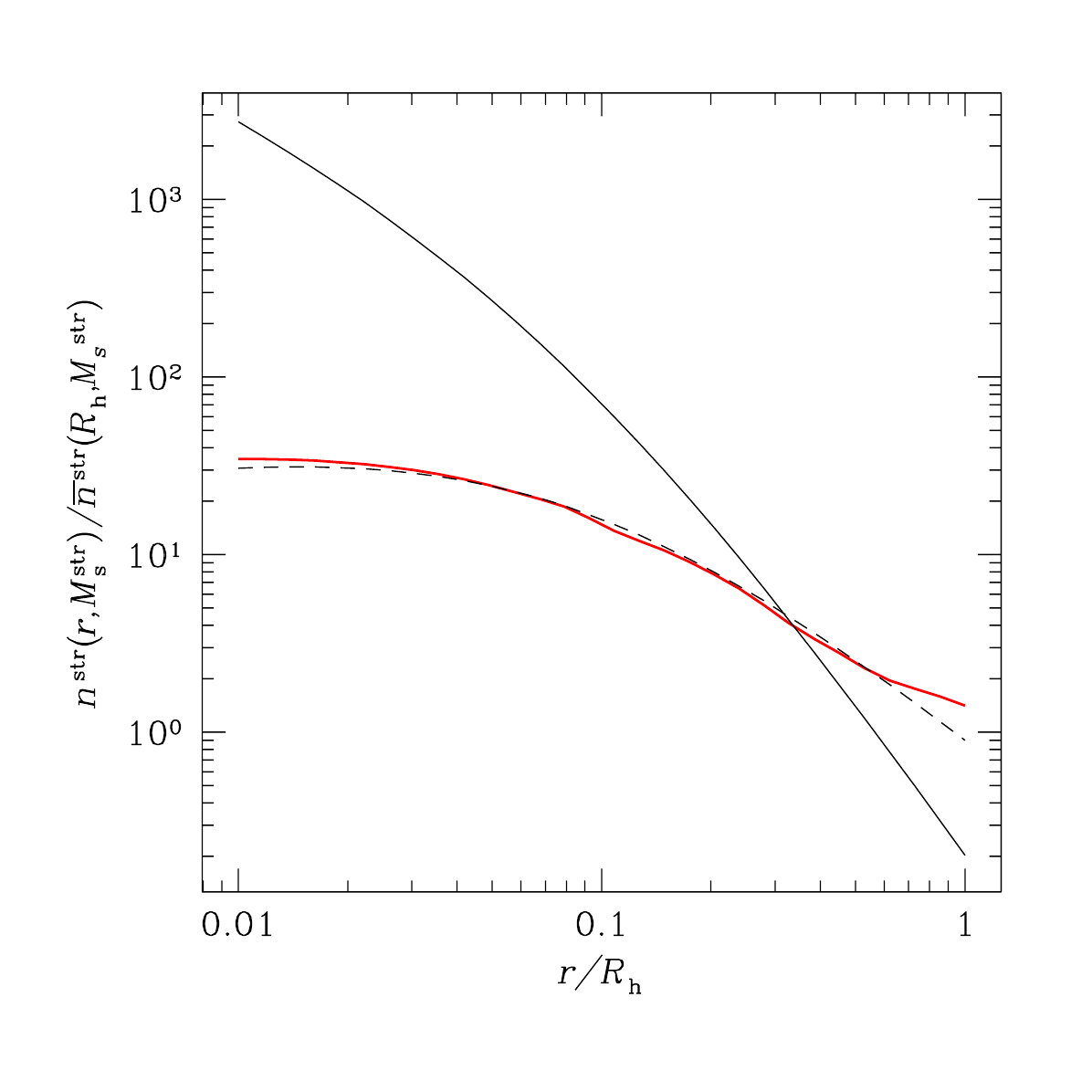}}
\caption{Scaled number density profiles of stripped subhalos predicted by CUSP (red line) for purely accreting MW-mass haloscompared to the fit by \citet{Han15} to the profile of the A halo in the Aquarius simulation \citep{Sea08} (black dashed line). The solid black line is the scaled halo density profile. (A colour version of this Figure is available in the online journal.)}
\label{f6}
\end{figure}

\begin{figure}
\centerline{\includegraphics[scale=.56,bb= 30 80 510 510]{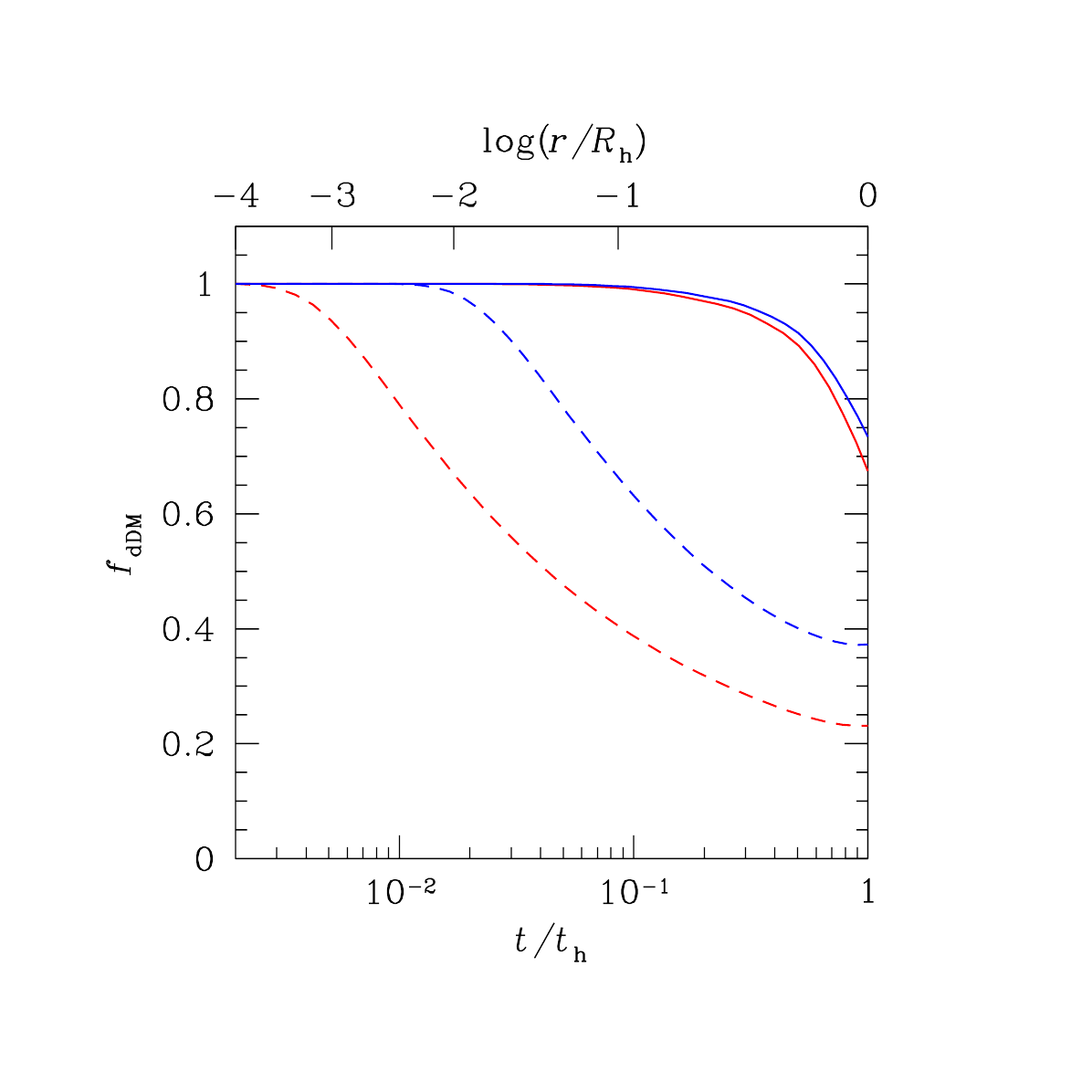}}
\caption{Total dDM mass fraction as a function of radius (upper
ticks) and cosmic time when the halo reached that radius (lower
ticks) predicted by CUSP for the same halos as in Figure \ref{f6} in a 100 GeV WIMP universe (solid red line) and an Aquarius-like \citep{Sea08} simulation (solid blue line). In dashed lines the corresponding mass fractions of accreted dDM. No results of simulations are available in this case.}
(A colour version of this Figure is available in the online journal.)
\label{f7}
\end{figure}

Equation (\ref{dens2}) shows that the mean spherically averaged scaled number density of stripped subhalos of different truncated masses $\clM\tr$ overlap in one only profile as a consequence of the separability of the abundance of accreted subhalos through equation (\ref{M}).
But this profile is bent with respect to $\rho\h(r)$  
because the dDM mass fraction increases towards the halo center due to the more marked stripping there (see Fig.~\ref{f6}). 

Taking into account the relation
\beq
1-f\dc\str(r\h)=[1+f\st(r\h)]\,\mu(r\h)[1-f\dc\acc(r\h)]
\label{fdDM}
\eeq
that follows from equations (\ref{rhod}) and (\ref{M}) and the separability of ${\cal N}\str(r\h,\clM\tr)$ and ${\cal N}\acc(r\h,\clM\tr)$, the integration over $r\h$ of equation (\ref{M}) leads to the relation
\beq
{\cal N}\str(\clM\tr)=[1-\bar f\dc(R\h)]{\cal N}\acc(\clM\tr)
\eeq
between the respective MFs. On the other hand, from the $f\dc\str(r\h)$ and $f\dc\acc(r\h)$ dDM mass fractions profiles (Fig.~\ref{f7}), we can compute the global total and accreted dDM mass fractions in halos. We find that about $\sim 90$ \% of the mass of current Milky Way-mass halos is in the form of dDM (Q xiii).

Notice that all halo profiles can be expressed in terms of $\rho\h(r)$, $f\dc\acc(r)$, $\mu(r)$ and $f\st(r)$. Since $\rho\h$ is roughly universal as a function of the scaled radius, $r/R\h$, and the $f\dc\acc$ profile following from $f\dc\acc(t)$ and $t(r/R\h)$ is also universal (see App.~\ref{approximation}), $\mu$ and $f\st$ are too \citep{Sea19b}. Thus, all properties regarding accreted and stripped subhalos in {\it purely accreting} halos (or having suffered the last major merger a very long time ago) are roughly universal (Q i). 

\section{Major Mergers and Gaussian Window}\label{mergers}

The imprints of tidal stripping on subhalos and dDM are never erased, so these components retain the memory of the halo past history. Thus, the abundance and radial distribution of stripped subhalos and the associated dDM in halos depend on the halo assembly history (see \citealt{Sea19c} for their derivation from those of purely accreting halos).

But {\it all the remaining halo properties}, including those regarding accreted subhalos and dDM, {\it are independent of the halo assembly history}, and can be obtained assuming PA (D vii). This surprising result, shown next, explains the empirical fact that all relaxed halos have similar properties regardless of the time of their last major merger (see also the discussion in Sec.~\ref{summ}).

In PA all halo properties evolve in a continuous manner. When a homeoid (or ellipsoidal shell) collapses, it crosses the homeoids having previously collapsed, with similar triaxial shapes and orientations. The energy outflow (in Lagrangian coordinates) set in such shell-crossings causes the particle apocenter loci to contract (and change their triaxial shape) until the correlation between the orbital phases of particles in neighboring shells is lost. Virialization is thus achieved through the randomization of orbital phases that goes together with the contraction of the system. However, the order of the particle apocenter loci is kept unaltered, implying that the radial mapping of the system is essentially conserved. It is thus unsurprising that the radial profiles and triaxial shape and kinematics of halos can be inferred from those of protohalos, and vice-versa. Consequently, there is no increase of entropy, meaning that the randomization of particle velocities through shell-crossing, which affects their orbital phases, {\it is not a real relaxation of the system}.

On the contrary, there is no continuity of halo properties interrupted in major mergers. Particles previously assembled in each of the progenitors and orbiting within it suddenly falls towards the new center of mass of the system where they mix up. This mixing yields a new phase of less symmetric shell-crossing around the center of mass and energy pumping outwards until the particle orbital phases are randomized, and the system stops contracting like in accretion. However, in this case, the particle apocenter order within each progenitor is not translated into the new structure, so there is a randomization of both particle phases and apocenters themselves, which causes the scrambling of particles in the new system. Such a scrambling destroys and smears out the progenitors so that the system becomes triaxial at all scales, from the largest one corresponding to the peak tracing the new halo, downwards. In other words, the memory of the initial multipolar mass distribution at small scales is lost, and the resulting mass distribution cannot be inverted anymore. Consequently, there is an increase of entropy, meaning that the randomization of particles velocities, which affects in this case the order of their apocentric radii and not just their orbital phases, {\it is a real (violent) relaxation of the system}. Shortly, the chaotic orbits followed by particles during the virialization of halos in major mergers cause the entropy increase in the two processes \citep{Bea19}.

Of course, the structural profiles (mass density, number density of accreted subhalos, and the accreted dDM density), kinematic profiles, and eccentricity profiles of merged halos can be inverted as if they had been set by PA. The result will be the structural profiles (mass density and number density of nested peaks) and eccentricity profiles of {\it fake} protohalos evolving into the merged halos by PA. The mean profiles of real merged halos are thus equal to the mean profiles arising from their respective fake protohalos. But the latter mean profiles coincide with the mean profiles of real halos evolving by PA from protohalos because the curvature of a peak does not depend on the small-scale mass distribution inside (and outside) the protohalo, so the mean curvature $\hat x$ of fake and real peaks evolving by PA with given density contrast and scale are the same. Thus the differential
equation (\ref{dmd}) governing the respective mean peak trajectories is
also the same, and, since the two kinds of peaks have identical characteristics at the scale $R$ and time $t$ of the final halos, the two mean $\delta(R)$ trajectories coincide, just as the respective mean $\ep(R)$ and $\es(R)$ trajectories following from them. Therefore, even though the small-scale mass distributions in protohalos of merged and purely accreting halos with given mass, triaxial shape, and kinematics at $t$ are very different, {\it their respective mean unconvolved profiles coincide.}

Moreover, not only do the mean profiles of both kinds of halos coincide, but the profiles of every individual purely accreting halo coincide with those of one merged halo, and vice-versa. Indeed, given a merged halo, we can reshuffle its particles in the protohalo, keeping the convolved density profile and triaxial shape of the peak unaltered. By doing this the real protohalo is converted into a fake protohalo evolving by PA into the merged halo. Conversely, given a purely accreting halo, we can concentrate its particles along the radial direction in a few small scale undulations, keeping the convolved density profile of the peak unaltered, then along their homeoids, keeping the triaxial shape of the peak unaltered, so as to build up a few massive clumps (i.e. peaks on smaller scales) within the large-scale peak without altering its convolved properties. This way the real protohalo of the purely accreting halo is converted into a protohalo of a merged halo with the same profiles as the original purely accreting halo. 

Thus both halo populations are indistinguishable, globally (the whole ensemble) as well as individually (every single realization). The only difference between them is in the entropy accompanying their collapse and virialization: in halos evolving by PA there is no entropy increase because the radial mapping of the protohalo is conserved, whereas in major mergers the entropy increases due to the reshuffling of particles produced in the merger at $t$, or in the protohalo at $\ti$ causing the new protohalo to evolve by PA to the same final halo. But the entropy increase in major mergers can only be detected provided the initial configuration is known. This is what happens in simulations where it is seen, indeed, that major mergers cause an entropy increase with respect to the case of accretion compatible with no increase \citep{Oea20}.

Thus halo properties do not depend on their assembly history (Q xiv). This result has in turn the following important consequence for the smoothing window. 

As the halo assembly process carries the loss of information on small scales of collapsing patches, the smoothing window that identifies those patches must filter out those small scales, i.e. it must be of compact support in Fourier space. Any other window with compact support in physical space as required to have finite collapsing patches, such as the top-hat one, has a Fourier transform with non-null values out to infinity. And any other window with compact support in Fourier space, such as the $k$-sharp window, does not encompass finite collapsing patches because its inverse Fourier transform has non-null values out to infinity. The way this problem is circumvented in the ES formalism is by using the $k$-sharp window with varying 0th-order spectral moment to identify the patches of different masses which should collapse according to the top-hat window with identical 0th-order spectral moments. Unfortunately, there is no reason for the masses seen by the two different windows with identical 0th-order spectral moment to coincide. 

Thus the Gaussian is the only smoothing window that accounts for the entropy increase in halo clustering (due to major mergers), and finds in a fully consistent manner the finite patches with known time of collapse and virialization (D iv). In this respect we remark that CUSP does provide the relation between the 0th-order spectral moments of the Gaussian and the top-hat windows that see collapsing patches of the same mass (eqs.~[\ref{sig}] and [\ref{sigma}]).

\section{Summary and Discussion}\label{summ}

CUSP is a rigorous formalism for the analytic treatment, from first principles and with no free parameters (but for the modeling of subhalo tidal heating), of halo clustering in hierarchical cosmologies. Taking advantage of the fact that the properties of halos do not depend on their assembly history, it uses the one-to-one correspondence between halos at $t$ and peaks with known properties in the smoothed density field at $\ti$ to infer the abundance and properties of the final objects by monitoring the monolithic ellipsoidal collapse and virialization (through shell-crossing) of their seeds.

The achievements of CUSP are multiple: it overcomes numerous fundamental difficulties (Ds) for the analytic follow up of DM clustering; it allows one to extend the results of simulations to any hierarchical cold or warm DM cosmology, any halo mass-definition, and any arbitrary halo radius, mass, and redshift; and it unravels the origin of all macroscopic halo properties and their characteristic features. In particular, it answers the following intriguing questions (Qs) risen by simulations:

\begin{itemize}

\item Q i) Why are the scaled spherically averaged profiles of halos so nearly universal?

\item Q ii) Why is the halo MF well predicted assuming monolithic collapse?

\item Q iii) Why does the halo multiplicity function privilege the FoF(0.2) halo-finding algorithm?

\item Q iv) What is the inner asymptotic slope of the halo density profile?

\item Q v) Why is the virial radius of a halo equal to that of a homogeneous spherical system having conserved the energy during collapse and virialization?

\item Q vi) What is the relation between the halo triaxial shape and the velocity anisotropy?

\item Q vii) Why is the pseudo phase-space density profile a power-law of index $-1.875$?

\item Q viii) Where does the universal anisotropy-density relation come from? 

\item Q ix) What are the typical halo inner and outer asymptotic ellipticity and prolateness?

\item Q x) Why is the subhalo cumulative MF close to a power-law of index $-1$?

\item Q xi) Why is the number density profile of accreted subhalos indepencdent of subhalo masses and nearly proportional to the halo density profile?

\item Q xii) Why is the truncated-to-original subhalo mass ratio independent of the subhalo mass?

\item Q xiii) How much diffuse DM is there in halos, and what is its spatial distribution?

\item Q xiv) Why are the properties of halos with very different assembly histories so similar?

\item Q xv) Where does the halo assembly bias come from?

\end{itemize}

Furthermore, CUSP has led to the following findings: there is a one-to-one halo (non-nested) peak connection, extensible to subhalos and nested peaks of any level; during accretion halos grow inside-out so that virialization preserves the radial mapping of the system; in major mergers halos are scrambled so that violent relaxation causes the memory loss of the system; the inner properties of halos (except for those related to stripping) do not allow one to distinguish whether and when halos have suffered major mergers; the total abundance of accreted and stripped subhalos coincides except for their normalization depending on the abundance of accreted and total dDM; the stripping-related properties of ordinary halos having suffered major mergers can be inferred from those derived assuming PA \citep{Sea19c}; and the filtering of the initial density field allowing one to monitor halo clustering must be carried out by means of a Gaussian window for consistency with the entropy increase produced in major mergers.

As mentioned, CUSP shows that the properties of halos do not depend on their assembly history. What is then the origin of the conflict between the results of simulations supporting this result \citep{Hea99,Hea06,WW09,Bea12}, and those indicating that halo properties depend on the frequency of major mergers, the so-called `assembly bias' \citep{Gea01,Gea02,ST04,FM09,FM10,Hea09,Gea19,Cea20,Hea20,Rea20,Wea20}. According to CUSP, what really causes halo properties to depend on the local density as found in the latter simulations is not that halos lying in different local densities suffer major mergers at a higher rate \citep{FM09}, but the different background densities of their seeds. Indeed, the local density of protohalos affects not only the frequency of major mergers, which as shown has no repercussion in the inner halo properties, but also and mainly their accretion rate \citep{FM10}, which sets their inner properties through the inside-out growth of purely accreting halos, and of halos suffering major mergers as well. 

In fact, if instead of monitoring the mean $\delta(S)$ trajectory for unconstrained peaks we had monitored that trajectory for peaks lying in backgrounds of different density contrasts $\delta'$ (e.g. \citealt{MSS95,Jea14b}) we would have been led to the mean profiles of halos {\it in the corresponding final local densities}. Clearly, these mean profiles would be somewhat different from those found for all halos regardless of their location. We thus see that the local density of halos affects their typical inner properties, even though major mergers leave no imprints on them (Q xv). 

Even though CUSP solves the main issues regarding DM halos it can still be improved. We are not thinking so much in minor effects such as the mass loss by halos in encounters and its eventual reaccretion \citep{vdB02,Wa11} or dynamical friction which alters the properties of substructure regarding massive subhalos, but in more fundamental effects such as the gravitational torque between neighbouring (proto)halos or the gravitational drag of baryons. The former is at the base of the small angular momentum and tidally-supported elongation of halos \citep{D70,W84,SS93}, and the latter yields notable deviations in halo structure from those of pure DM objects derived here \citep{GO99,GZ02,Gea12,PG12}.

\begin{acknowledgments}
Funding for this work was provided by the Spanish MINECO under projects CEX2019-000918-M of ICCUB (Unidad de Excelencia `Mar\'ia de Maeztu') and PID2019-109361GB-100 (co-funded with FEDER funds) and the Catalan DEC grant 2017SGR643. We thank Prof. Ravi K. Sheth for fruitful comments.
\end{acknowledgments}

\begin{appendix}

\section{Analytic Approximation for the Halo Density Profile}
\label{approximation}

Taking into account that, in limited mass ranges, the power spectrum is well-approximated by a power-law $P(k)=C k^n$ with effective index $n$, the j-th spectral moment on scale $R_{\rm f}$ for any window f with Fourier transform $W_f(kR_{\rm f})$ 
takes the form
\beq
(\sigma_{\rm j}^{\rm f})^2(R_{\rm f})\approx \frac{C}{2\pi^2 R_{\rm f}^{n+3+2{\rm j}}}\int_0^\infty \der x\, x^{n+2(1+{\rm j})}\, W_{\rm f}^2(x)\,.
\eeq
Comparing the expression arising from the Gaussian and the top-hat windows we arrive at (eq.~[\ref{rm}])
\beq
r_{\rm R}(M,t)\approx\left[Q_{\rm j}\,\frac{\sigma_{\rm j}\F[R\F(M),\ti]}{\sigma_{\rm j}[R(M,t),\ti]}\right]^{\frac{2}{n+3+2{\rm j}}}\,~~~~~~~~~~~~~~~~~~~~~~~~~~~~~~~~~~~~~Q_{\rm j}^2\equiv  \frac{\int_0^\infty \der x \, x^{n+2(1+{\rm j})}\,W_{\rm G}^2(x)}{\int_0^\infty \der x \, x^{n+2(1+{\rm j})}\,W^2_{\rm th}(x)}\,,
\label{rsigmaj}
\eeq
for any j, where $W_{\rm G}$ and $W_{\rm th}$ are the Fourier transforms of the Gaussian and top-hat windows, respectively. For j$=0$, we have (eqs.~[\ref{sigma}]-[\ref{St}])
\beq
r_{\rm R}(M,t)\approx
\left\{Q_0\left[1+S(t)\frac{a(t)}{D(t)}\nu\F(M,t)\right]^{-1}\right\}^{\frac{2}{n+3}}~~~~~~~~~~~~~~~~~~~~~~~S(t)=s_0[1+s\,a(t)]\,,
\eeq
where we have used $d\approx 1$ and $[a(t)/D(t)]\nu\F(M,t)\la 1$ (see Table 1).\footnote{Indeed, the quantity $[a(t)/D(t)]\nu\F(M,t)$ is approximately equal to $\nu(M,t)$ whose typical values are shown in Fig.~\ref{f1}.} Taking into account that $s_0\sim 10^{-2}$, $s\,a(t)\la 1$, we have $r_\sigma\approx 1$ (eq.~[\ref{sigma}]), $r_{\rm R}(M,t)\approx Q_0^{\frac{2}{n+3}}$, and (eq.~[\ref{rmF}])
\beq
R(M,\ti)\approx Q_0^{\frac{2}{n+3}}\left[\frac{3M}{4\pi \rho\cc(\ti)}\right]^{1/3}\,.
\label{rm2}
\eeq
The values of $n$ and $Q_0$ vary with mass range, cosmology, and halo mass definition, although in all cases of interest they are about $-1.5$ and $0.5$, respectively.

Taking into account the relation (\ref{rm2}), the peak trajectory solution of the differential equation (\ref{dmd}) with $\hat x(R,\delta)\approx \gamma \nu$ replacing $x(R,\delta)$ and $\delta(t)$ given by equations (\ref{deltat})--(\ref{cc}) then leads to the following universal mean halo mass accretion history (MAH)
\beq
\frac{M\h(t)}{M\h(t\h)}\equiv \frac{M\{R[\delta(t)],t\h\}}{M\h(t\h)} \approx \left[\frac{\delta\cc\F(t)}{\delta\cc\F(t\h)}
\frac{a(t)D^2(t\h)}{D^2(t)}\right]^{3/m},
\label{Mt}
\eeq
with $m=-[(n+3)/2]^{3/2}$, as found in simulations (e.g. \citealt{LEtal13,Cea15}). Together with the definition of the virial radius $r\h$ of the halo at $t$ scaled to its value $R\h=r\h(t\h)$ at $t\h$,
\beq
\frac{r\h(t)}{R\h}=\left[\frac{M\h(t)\Delta\vir(t\h)\rho\cc(t\h)}{M\h(t\h) \Delta\vir(t)\rho\cc(t)}\right]^{1/3},
\label{rt}
\eeq
equation (\ref{Mt}) yields an approximate parametric expression for the mean halo mass profile $M\h(r\h/R\h)$ according to the SO($\Delta\vir$) halo mass definition, directly arising from the mean peak trajectory $\delta(R)$ (see e.g. \citealt{H00} for the analytic expressions of $\delta\cc\F(t)$, $\Delta\vir(t)$, and $D(t)$).\footnote{Alternatively we can plug the function $t(M\h)$ inverse of equation (\ref{Mt}) into equation (\ref{rt}) to directly obtain the mass profile $M\h(r\h)$ of the halo with mass $M\h(t\h)$ at $t\h$.} Then, the relation
\beq 
\rho\h\left(\frac{r\h}{R\h}\right)=\frac{\Delta\vir(t\h)\rho\cc(t\h)}{4\pi}\,\frac{\der [M\h(t)/M\h(t\h)]}{\der t}\left\{\frac{\der [r\h(t)/R\h]^3}{\der t}\right\}^{-1},
\eeq
following from the inside-out growth of accreting halos, together with the time-derivatives of the relations (\ref{Mt}) and (\ref{rt}) sets a parametric expression for the closely universal mean spherically averaged halo density profile.

\section{Spherical Quantities}\label{sphericalApp}

Given any arbitrary mass distribution, the density and potential at a radius {\bf r} from the center of mass of the system can be split as the sum of their spherical average plus a residual,
\beq
\,\rho({\bf r})=\rho(r)+\delta\rho({\bf r})~~~~~~~~~~~~~~~~~~~~~~
\label{split1}
\Phi({\bf r}) =\Phi(r)+\delta\Phi({\bf r})\,.
\eeq
In {\it triaxial systems}, the variances $\lav\delta\rho^2\rav(r)/\rho^2(r)$, $\lav\delta\Phi^2\rav(r)/\Phi^2(r)$ and covariance $|
\lav\delta\rho\delta \Phi\rav(r)/[\rho(r)\Phi(r)]|$ over spherical surfaces of radius $r$ are always less than one \citep{Sea12b}. From now on, angular brackets mean spherical average. 

Given those decompositions, any quantity $F(r)$ inside $r$ (or over the sphere of radius $r$) can also be split as the sum of a function ${\cal F}(r)$, with the same meaning as $F(r)$, but holding for the spherically symmetrized system, hereafter simply the `the spherical quantity', plus a residual $\delta{\cal F}(r)$, with $\delta{\cal F}(r)/{\cal F}(r)< 1$ \citep{Sea12b}.

In particular, we have $M(r)={\cal M}(r)+\delta{\cal M}(r)$, being
\beq
{\cal M}(r)=4\pi\int_0^{r} \der x\, x^2\,\rho(x)\,~~~~~~~~~~~~~~~~~~~~~~~~~~~~~~~~~~~~~~~~~~~~~~~~~~~~~~\delta{\cal M}(r)=0,
\label{mass}
\eeq
and $E(r)={\cal E}(r)+\delta{\cal E}(r)$, being 
\beq 
{\cal E}(r)= 4\pi \int_0^{r} \der
x\,x^2\,\rho(x)\left[\frac{{\cal s}^2(x)}{2}-\frac{G{\cal M}(x)}{x}\right]\,~~~~~~~~~~~~~~~~~~~~~
\delta{\cal E}(r)=2\pi\!\int_0^r \der
x\,x^2\frac{\delta {\cal s}^2(x)}{2}+
  \lav\delta\rho\,\delta\Phi\rav(x)\,,
\label{energy}
\eeq
where ${\cal s}^2(r)$ is the spherical velocity variance, related to the real velocity variance through $\sigma^2(r)={\cal s}^2(r)+\delta {\cal s}^2(r)$ \citep{Sea12b}. Notice that ${\cal E}(r)$ accounts for the energy lost by the system through shell-crossing (if any), but not for the potential energy exchange between the sphere and the rest of the system due to its asphericity, which is included in the residual $\delta {\cal E}(r)$.

If the system is, in addition, {\it in equilibrium}, the steady collisionless Boltzmann equation leads to the virial relation \citep{Sea12b}
\beq 
\frac{2{\cal E}(r)}{{\cal W}(r)}=1-{\cal S}(r)\,,
\label{vir0l}
\eeq
in terms of the spherical total energy ${\cal E}$, spherical potential energy ${\cal W}$, and spherical surface term ${\cal S}$, with the two latter respectively equal to
\beq
{\cal W}(r)=-4\pi\int_0^r \der x\,x^2\,\rho(x)\,\frac{G{\cal M}(x)}{x}\,,~~~~~~~~~~~~~~~~~~~~~~~~~~~~~~~~~~{\cal S}(r)=\frac{4\pi r^3\rho(r)\, {\cal s}^2\rad(r)}{{\cal W}(r)}\,,
\label{spot2}
\eeq
where ${\cal s}^2\rad(r)$ is the spherical radial velocity variance, related to the ordinary one through $\sigma\rad^2(r)={\cal s}^2\rad(r)+\delta {\cal s}^2\rad(r)$, being
\beq 
\delta {\cal s}^2\rad(r)= \frac{1}{r^3\rho(r)}\int_0^r \der x\,
x^2 \left\{\delta {\cal s}^2(x)-x \lav\delta\rho\, \derpr \delta\Phi\rav (x)\right\}\,.
\label{dsr}
\eeq

Differentiating equations (\ref{mass}), (\ref{energy}) and (\ref{vir0l}), we arrive at the relations
\beq
\rho(r)= \frac{1}{4\pi r^2}\frac{\der {\cal M}}{\der r}\,,
\label{rhot}
\eeq
\beq 
{\cal s}^2(r)=2\left[\frac{\der {\cal E}/\der
r} {\der {\cal M}/\der r}+ \frac{G{\cal M}(r)}{r}\right]~~~~~~~~~~~~~~~~~~~~~~~~{\rm and}~~~~~~~~~~~~~~~~~~~~~~~~{\cal s}\rad^2(r)=\frac{2{\cal E}(r)-{\cal W}(r)}{r\,\der {\cal M}/\der r}\,.
\label{sig2}
\eeq
These relations are independent of the asphericity of the system, so, contrarily to the $\sigma^2(r)$ and $\sigma^2\rad(r)$ profiles which involve the quantities $\delta{\cal s}^2(r)$ and $\delta{\cal s}\rad^2(r)$ (eqs.~[\ref{energy}] and [\ref{dsr}]), the $M(r)$, $\rho(r)$, ${\cal s}^2(r)$, and ${\cal s}^2\rad(r)$ profiles do not depend on the shape of the system. Thus, {\it as long as we are not concerned with the former quantities but just with the later ones, we can assume spherical symmetry}. This concerns the expressions for both the halo and the protohalo as well as the relations between the two objects.

It is worthwhile mentioning that, since in ideal spherically symmetric halos particle orbits are purely radial (particles collapse and bounce radially), we have $\sigma^2\h(r\h)=(\sigma^2\rad)\h(r\h)$. Equations (\ref{sig2}) then lead to a differential equation for ${\cal E}\h(r\h)$ that can be readily integrated for the boundary condition ${\cal E}\h(0)=0$. The result is
\beq 
{\cal E}\h(r\h)=-R\int_0^{r\h} \der x\left[4\pi\,\rho\h(x)\,G{\cal M\h}(x)+\frac{{\cal W}\h(x)}{2x^2}\right]\,.
\label{diff}
\eeq
From equations (\ref{diff}) and (\ref{energy}), we can compute the dissipation factor by shell-crossing, $D(M\h)\equiv  E\h(M\h)/ E\p(M\p=M\h)$. The result is such that $\rho\h(r\h)$ is necessarily outwards decreasing \citep{Sea12a}.

\section{Eccentricities}
\label{eccentricity}

Given a triaxial system, the homeoid with isodensity $\rho\el$ and semiaxes $a_1\ge a_2\ge a_3$, can be labelled by means of the radius $r$ defined as
\beq
r=\left[\frac{1}{3}\left(a_1^2+a_2^2+a_3^2\right)\right]^{1/2}\,.
\label{R1}
\eeq
The density at the point ${\bf r}=\sphc$ over the homeoid then takes the form
\beq
\rho({\bf r})=\rho\el(r)\left[1-\frac{\ep^2(r)+\es^2(r)}{3}\right]+\left[\sin^2\theta \cos^2\phi+\frac{\sin^2\theta \sin^2\phi}{1-\esp^2(r)}+
\frac{\cos^2\theta}{1-\eps^2(r)}\right]\,,
\label{ellips}
\eeq
where the 
$z$ axis is taken aligned with the major axis of the homeoid,
and, depending on the orientation of the $x$ and $y$
axes relative to the minor and intermediate ellipsoid semiaxes, the primary or secondary eccentricities, $\ep$ and $\es$, are defined as
\beq
\ep=\left(1-\frac{a_3^2}{a_1^2}\right)^{1/2}~~~~~~~~~~~~~~~~~~~~~~~~~~~~~~~{\rm and}~~~~~~~~~~~~~~~~~~~~~~~~~~~~~~\es=\left(1-\frac{a_2^2}{a_1^2}\right)^{1/2},
\label{primary}
\eeq
or vice-versa. Thus the spherically averaged density at $r$ is 
\beq
\rho(r)=\frac{\rho\el(r)}{3}\left[1-\frac{\ep^2(r)+\es^2(r)}{3}\right]G(r),~~~~~~~~~~~~~~~~~~~~~~~~~~~~
G(r)=a_1^2(r)\left[\frac{1}{a_1^2(r)}+\frac{1}{a_2^2(r)}+\frac{1}{a_3^2(r)}\right]\,.
\label{ellipsbis}
\eeq

Dividing equation (\ref{ellips}) by equation (\ref{ellipsbis}), we find
\beq
1\!+\!\frac{\delta\rho({\bf r})}{\rho(r)}\!=\!\frac{3}{G(r)}\!
\left[\sin^2\!\theta \cos^2\!\phi+\frac{\sin^2\!\theta \sin^2\!\phi}{1\!-\!\esp^2(r)}+
\frac{\cos^2\!\theta}{1\!-\!\eps^2(r)}\right],
\label{ellips2}
\eeq
and the definitions (\ref{primary}) then lead to a scaled density variance over the sphere of radius $r$ given by
\beq
\frac{\lav\delta\rho\rav^{\!2}(r)}{\rho^2(r)}=-\frac{2}{5}\bigg\{1
-\frac{3[(1-\ep^2)^2(1-\es^2)^2+(1-\ep^2)^2+(1-\es^2)^2]}
{[(1-\ep^2)(1-\es^2)+(1-\ep^2)+(1-\es^2)]^2}\bigg\}(r).\label{4th}
\eeq

On the other hand, the Poisson equation relates $\delta \Phi({\bf r})/\Phi(r)$ to $\delta \rho({\bf r})/\rho(r)$, and, integrating over the solid angle, we can express the scaled potential variance and density-potential covariance profiles in terms of the scaled density variance, and, through equation (\ref{4th}), of the eccentricity profiles \citep{Sea12b}.\\

The rest of this Appendix is devoted to relate the eccentricities of the halo and the protohalo taking into account the two conservation conditions satisfied during ellipsoidal collapse given in Section \ref{shape}.

Taking into account the relations (\ref{primary}), equation (\ref{emass}) takes the form
\beq
\frac{[(1-\ep\h^2)(1-\es\h^2)]^{1/3}}{1+(1-\ep\h^2)+(1-\es\h^2)}(r\h)=
\frac{[(1-\ep\p^2)(1-\es\p^2)]^{1/3}}{1+(1-\ep\p^2)+(1-\es\p^2)}(r\p).
\label{6th}
\eeq

On the other hand, replacing the expressions of $\delta{\cal E}$ (eq.~[\ref{energy}]) in the halo and protohalo into equation (\ref{cale}), and differentiating it, we are led to
\beq
\frac{1}{{\cal D}(r\h)}\!\left[\frac{\lav\delta\rho\h\delta\Phi\h\rav(r\h)}{\rho\h(r\h)}+\delta\sigma^2\h(r\h)\right]
=\frac{\lav\delta\rho\p\delta\Phi\p\rav(r\p)}{\rho\p(r\p)}
\label{1st}~~~~~~~{\cal D}(r\h)=\frac{10r\h^2}{3{\cal M}^2(r\h)} \int_0^{r\h} \der x \left[4\pi \rho(x) {\cal M}(x)+\frac{{\cal W}(r)}{2 x^2}\right]\!.
\eeq
And, taking the density-potential covariance as a function of density variance, and the relation between this latter and the eccentricities (App.~\ref{eccentricity}), we arrive at
\beq
U(r_{\rm i})Z\h(r\h)-\frac{3[(1\!-\!\ep\h^2)^2(1\!-\!\es\h^2)^2+(1\!-\!\ep\h^2)^2+(1\!-\!\es\h^2)^2]}
{[(1\!-\!\ep\h^2)(1\!-\!\es\h^2)+(1\!-\!\ep\h^2)+(1\!-\!\es\h^2)]^2}(r\h)\!=\!1\!-\!
\frac{3[(1\!-\!\ep\p^2)^2(1\!-\!\es\p^2)^2+(1\!-\!\ep\p^2)^2+(1\!-\!\es\p^2)^2]}
{[(1\!-\!\ep\p^2)(1\!-\!\es\p^2)+(1\!-\!\ep\p^2)+(1\!-\!\es\p^2)]^2}(r\p)
\label{7th}
\eeq
where
\beq
U(r\h,r\p)\equiv \frac{1}{{\cal D}(r\h)}\frac{\Phi\h(r\h)V\p(r\p)}{\Phi\p(r\p)V\h(r\h)}\,~~~~~~~~~~~~~~~{\rm and}~~~~~~~~~~~~~~~~
Z\h(r\h)\equiv 1- \frac{5}{2}\frac{\sigma\h^2(r\h)-{\cal s}\h^2(r\h)}{\Phi\h(r\h)}V\h(r\h)\,,
\label{S}
\eeq 
with $V(r)=1-\xi(r)\gamma(r)\left\{1-[1+\kappa(r)]\gamma(r)-\der\ln \gamma/\der\ln r\right\}$, and $\xi(r)$, $\kappa(r)$, and $\gamma(r)$ equal to the DM two-point correlation function and the logarithmic derivatives of the density variance and the density-potential covariance, respectively. At small radii $Z\h(r)-1$ is negligible ($|\Phi\h(r\h)|\gg \sigma\h^2(r\h)\sim GM\h(r\h)/r\h$) so, for $U(r\h)=1$, equation (\ref{7th}) becomes an identity relation like equation (\ref{6th}). In fact, the algebraic equations (\ref{6th}) and (\ref{7th}) are solvable only for a very narrow range of $U(r)$ values around unity (\citealt{Sea12b}).

\end{appendix}

\end{document}